\documentclass[12pt]{article}
\usepackage{graphics,graphicx}
\usepackage{amssymb,epsfig,amsmath,euscript,array}
\usepackage{cite}
\usepackage{axodraw}
\usepackage{pstricks}
\usepackage{color}

\makeatletter
\@addtoreset{equation}{section}
\makeatother
\renewcommand{\theequation}{\thesection.\arabic{equation}}

\newcommand{\startappendix}{
\setcounter{section}{0}
\renewcommand{\thesection}{\Alph{section}}
\renewcommand{\theequation}{\Alph{section}.\arabic{equation}}}

\newcounter{multieqs}

\newcommand{\be}{\begin{equation}}
\newcommand{\ee}{\end{equation}}

\newcommand{\bm}[1]{\mbox{\boldmath $#1$}}

\newcommand{\kslash}{k \!\!\! / }

\newcommand{\lslash}{l \!\! / }
\newcommand{\Pslash}{P \!\!\!\! / }

\newcommand{\islash}{i \!\!\! / }
\newcommand{\jslash}{j \!\!\! / }
\newcommand{\aslash}{a \!\!\! / }
\newcommand{\bslash}{{b \hspace{-6pt} \slash} }

\newcommand{\onslash}{1 \!\!\! / }
\newcommand{\twslash}{2 \!\!\!/ }
\newcommand{\thslash}{3 \!\!\!/ }
\newcommand{\foslash}{4 \!\!\! / }
\newcommand{\fislash}{5 \!\!\! / }

\newcommand{\mslash}{m \!\!\! / }

\def\bd{\begin{document}}
\def\ed{\end{document}}
\def\nn{\nonumber}
\def\bea{\begin{eqnarray}}
\def\eea{\end{eqnarray}}

\def\red{\color{red}}
\def\black{\color{black}}
\def\blue{\color{blue}}
\def\orange{\color{orange}}

\def\ab{(ijab)}
\def\ba{(ijba)}
\def\ijab{{\tr}_{+}(\islash\, \jslash\, \aslash \, \bslash)}
\def\ijba{{\tr}_{+}(\islash\, \jslash\, \bslash \, \aslash)}
\def\ijaP{{\tr}_{+}(\islash\, \jslash\, \aslash \, \Pslash)}
\def\ijPLa{{\tr}_{+}(\islash\, \jslash\, \Pslash_L \, \aslash)}
\def\ijaPL{{\tr}_{+}(\islash\, \jslash\, \aslash \, \Pslash_L)}
\def\ijPLza{{\tr}_{+}(\islash\, \jslash\, \Pslash_{L;z} \, \aslash)}
\def\ijaPLz{{\tr}_{+}(\islash\, \jslash\, \aslash \, \Pslash_{L;z})}
\def\ijPa{{\tr}_{+}(\islash\, \jslash\, \Pslash \, \aslash)}
\def\iaPb{{\tr}_{+}(\islash\, \aslash\, \Pslash \, \bslash)}
\def\ibPa{{\tr}_{+}(\islash\, \bslash\, \Pslash \, \aslash)}
\def\ijPmu{{\tr}_{+}(\islash\, \jslash\, \Pslash \, \mu)}
\def\ibmuP{{\tr}_{+}(\islash\, \bslash\, \mu \, \Pslash)}
\def\ibmua{{\tr}_{+}(\islash\, \bslash\, \mu \, \aslash)}
\def\iamub{{\tr}_{+}(\islash\, \aslash\, \mu \, \bslash)}
\def\jaPb{{\tr}_{+}(\jslash\, \aslash\, \Pslash \, \bslash)}
\def\ijmuP{{\tr}_{+}(\islash\, \jslash\, \mu \, \Pslash)}
\def\ijmum{{\tr}_{+}(\islash\, \jslash\, \mu \, \mslash)}
\def\ijmmu{{\tr}_{+}(\islash\, \jslash\, \mslash \, \mu)}
\def\ijmP{{\tr}_{+}(\islash\, \jslash\, \mslash \, \Pslash)}
\def\iabP{{\tr}_{+}(\islash\, \aslash\, \bslash \, \Pslash)}
\def\ijbP{{\tr}_{+}(\islash\, \jslash\, \bslash \, \Pslash)}
\def\jbPa{{\tr}_{+}(\jslash\, \bslash\, \Pslash \, \aslash)}
\def\ijPb{{\tr}_{+}(\islash\, \jslash\, \Pslash \, \bslash)}
\def\jbmua{{\tr}_{+}(\jslash\, \bslash\, \mu \, \aslash)}

\def\loablt{ {\tr}_{+}(\lslash_1\, \aslash \, \bslash\, \lslash_2)}

\def\ijlolt{{\tr}_{+}(\islash\, \jslash\, \lslash_1 \, \lslash_2)}
\def\ijltlo{{\tr}_{+}(\islash\, \jslash\, \lslash_2 \, \lslash_1)}
\def\ibloa{{\tr}_{+}(\islash\, \bslash\, \lslash_1 \, \aslash)}
\def\jaltb{{\tr}_{+}(\jslash\, \aslash\, \lslash_2 \, \bslash)}
\def\ialtb{{\tr}_{+}(\islash\, \aslash\, \lslash_2 \, \bslash)}
\def\bltloa{{\tr}_{+}(\bslash\, \lslash_2\, \lslash_1 \, \aslash)}
\def\jbloa{{\tr}_{+}(\jslash\, \bslash\, \lslash_1 \, \aslash)}
\def\ibPb{{\tr}_{+}(\islash\, \bslash\, \Pslash \, \bslash)}
\def\ijltb{{\tr}_{+}(\islash\, \jslash\, \lslash_2 \, \bslash)}

\def\ijloa{{\tr}_{+}(\islash\, \jslash\,  \lslash_1 \, \aslash)}
\def\ijblt{{\tr}_{+}(\islash\, \jslash\,  \bslash \, \lslash_2)}

\def\jakb{{\tr}_{+}(\jslash\, \aslash\, \kslash \, \bslash)}
\def\iakb{{\tr}_{+}(\islash\, \aslash\, \kslash \, \bslash)}

\def\tofo{{\tr}_{+}(\onslash\, \thslash\, \twslash \, \foslash)}
\def\foto{{\tr}_{+}(\onslash\, \thslash\, \foslash \, \twslash)}
\def\tofi{{\tr}_{+}(\onslash\, \thslash\, \twslash \, \fislash)}
\def\fito{{\tr}_{+}(\onslash\, \thslash\, \fislash \, \twslash)}

\def\lrangle#1#2{\langle #1\,#2\rangle}

\def\Li{{$\rm Li}_2$}
\def\eps{\epsilon}
\def\epsuv{{\epsilon_{\scriptscriptstyle{\mathrm{UV}}}}}

\let\bm=\bibitem
\let\la=\label

\def\npb#1#2#3{Nucl. Phys. {\bf{B#1}} #3 (#2)}
\def\plb#1#2#3{Phys. Lett. {\bf{#1B}} #3 (#2)}
\def\prl#1#2#3{Phys. Rev. Lett. {\bf{#1}} #3 (#2)}
\def\prd#1#2#3{Phys. Rev. {D \bf{#1}} #3 (#2)}
\def\cmp#1#2#3{Comm. Math. Phys. {\bf{#1}} #3 (#2)}
\def\cqg#1#2#3{Class. Quantum Grav. {\bf{#1}} #3 (#2)}
\def\nppsa#1#2#3{Nucl. Phys. B (Proc. Suppl.) {\bf{#1A}}#3 (#2)}
\def\ap#1#2#3{Ann. of Phys. {\bf{#1}} #3 (#2)}
\def\ijmp#1#2#3{Int. J. Mod. Phys. {\bf{A#1}} #3 (#2)}
\def\rmp#1#2#3{Rev. Mod. Phys. {\bf{#1}} #3 (#2)}
\def\mpla#1#2#3{Mod. Phys. Lett. {\bf A#1} #3 (#2)}
\def\jhep#1#2#3{J. High Energy Phys. {\bf #1} #3 (#2)}
\def\atmp#1#2#3{Adv. Theor. Math. Phys. {\bf #1} #3 (#2)}
\newcommand{\EQ}[1]{\begin{equation} #1 \end{equation}}
\newcommand{\AL}[1]{\begin{subequations}\begin{align} #1 \end{align}\end{subequations}}
\newcommand{\SP}[1]{\begin{equation}\begin{split} #1 \end{split}\end{equation}}
\newcommand{\ALAT}[2]{\begin{subequations}\begin{alignat}{#1} #2 \end{alignat}
                        \end{subequations}}
\def\beqa{\begin{eqnarray}}
\def\eeqa{\end{eqnarray}}
\def\beq{\begin{equation}}
\def\eeq{\end{equation}}
\def\sst{\scriptscriptstyle}
\def\thetabar{\bar\theta}
\def\Tr{{\rm Tr}}
\def\one{\mbox{1 \kern-.59em {\rm l}}}
 \def\Nh{\hat{N}}

\def\a{\alpha}      \def\da{{\dot\alpha}}
\def\b{\beta}       \def\db{{\dot\beta}}
\def\c{\gamma}  \def\G{\Gamma}  \def\cdt{\dot\gamma}
\def\d{\delta}  \def\D{\Delta}  \def\ddt{\dot\delta}
\def\e{\epsilon}        \def\vare{\varepsilon}
\def\f{\phi}    \def\F{\Phi}    \def\vvf{\f}
\def\h{\eta}
\def\k{\kappa}
\def\l{\lambda} \def\L{\Lambda}
\def\m{\mu} \def\n{\nu}
\def\o{\omega}
\def\p{\pi} \def\P{\Pi}
\def\r{\rho}
\def\s{\sigma}  \def\S{\Sigma}
\def\t{\tau}
\def\th{\theta} \def\Th{\Theta} \def\vth{\vartheta}
\def\X{\Xeta}
\def\z{\zeta}
\def\de{\partial}

\def\cA{{\cal A}} \def\cB{{\cal B}} \def\cC{{\cal C}}
\def\cD{{\cal D}} \def\cE{{\cal E}} \def\cF{{\cal F}}
\def\cG{{\cal G}} \def\cH{{\cal H}} \def\cI{{\cal I}}
\def\cJ{{\cal J}} \def\cK{{\cal K}} \def\cL{{\cal L}}
\def\cM{{\cal M}} \def\cN{{\cal N}} \def\cO{{\cal O}}
\def\cP{{\cal P}} \def\cQ{{\cal Q}} \def\cR{{\cal R}}
\def\cS{{\cal S}} \def\cT{{\cal T}} \def\cU{{\cal U}}
\def\cV{{\cal V}} \def\cW{{\cal W}} \def\cX{{\cal X}}
\def\cY{{\cal Y}} \def\cZ{{\cal Z}}

\def\ua{\underline{\alpha}}
\def\ub{\underline{\phantom{\alpha}}\!\!\!\beta}
\def\uc{\underline{\phantom{\alpha}}\!\!\!\gamma}
\def\um{\underline{\mu}}
\def\ud{\underline\delta}
\def\ue{\underline\epsilon}
\def\una{\underline a}\def\unA{\underline A}
\def\unb{\underline b}\def\unB{\underline B}
\def\unc{\underline c}\def\unC{\underline C}
\def\und{\underline d}\def\unD{\underline D}
\def\une{\underline e}\def\unE{\underline E}
\def\unf{\underline{\phantom{e}}\!\!\!\! f}\def\unF{\underline F}
\def\unm{\underline m}\def\unM{\underline M}
\def\unn{\underline n}\def\unN{\underline N}
\def\unp{\underline{\phantom{a}}\!\!\! p}\def\unP{\underline P}
\def\unq{\underline{\phantom{a}}\!\!\! q}
\def\unQ{\underline{\phantom{A}}\!\!\!\! Q}
\def\unH{\underline{H}}

\def\As {{A \hspace{-6.4pt} \slash}\;}
\def\bs {{b \hspace{-6.4pt} \slash}\;}
\def\Ds {{D \hspace{-6.4pt} \slash}\;}
\def\ds {{\del \hspace{-6.4pt} \slash}\;}
\def\ss {{\s \hspace{-6.4pt} \slash}\;}
\def\ks {{ k \hspace{-6.4pt} \slash}\;}
\def\ps {{p \hspace{-6.4pt} \slash}\;}
\def\pas {{{p_1} \hspace{-6.4pt} \slash}\;}
\def\pbs {{{p_2} \hspace{-6.4pt} \slash}\;}
\def\Ps {{P \hspace{-6.4pt} \slash}\;}
\def\Qs {{Q \hspace{-6.4pt} \slash}\;}

\def\Fh{\hat{F}}
\def\Vh{\hat{V}}
\def\Xh{\hat{X}}
\def\ah{\hat{a}}
\def\xh{\hat{x}}
\def\yh{\hat{y}}
\def\ph{\hat{p}}
\def\xih{\hat{\xi}}
\def\psit{\tilde{\psi}}
\def\Psit{\tilde{\Psi}}
\def\tht{\tilde{\th}}
\def\lt{\tilde{\lambda}}
\def\hl{\hat{\lambda}}
\def\hlt{\hat{\tilde{\lambda}}}
\def\llt{\tilde{l}}
\def\At{\tilde{A}}
\def\Qt{\tilde{Q}}
\def\Rt{\tilde{R}}
\def\Nt{\tilde{N}}

\def\at{\tilde{a}}
\def\st{\tilde{s}}
\def\ft{\tilde{f}}
\def\pt{\tilde{p}}
\def\qt{\tilde{q}}
\def\vt{\tilde{v}}
\def\nt{\tilde{n}}

\def\delb{\bar{\partial}}
\def\bz{\bar{z}}
\def\bD{\bar{D}}
\def\bB{\bar{B}}

\def\bk{{\bf k}}
\def\bl{{\bf l}}
\def\bp{{\bf p}}
\def\bq{{\bf q}}
\def\br{{\bf r}}
\def\bx{{\bf x}}
\def\by{{\bf y}}
\def\bR{{\bf R}}
\def\bV{{\bf V}}

\def\d{\delta}\def\D{\Delta}\def\ddt{\dot\delta}
\def\pa{\partial} \def\del{\partial}
\def\xx{\times}
\def\uno{\mbox{1 \kern-.59em {\rm l}}}
\def\trp{^{\top}}
\def\inv{^{-1}}
\def\dag{{^{\dagger}}}
\def\pr{^{\prime}}
\def\lan{\langle}
\def\ran{\rangle}
\def\rar{\rightarrow}
\def\lar{\leftarrow}
\def\lrar{\leftrightarrow}
\newcommand{\0}{\,\!}      
\def\one{1\!\!1\,\,}
\def\im{\imath}
\def\jm{\jmath}
\newcommand{\tr}{\mbox{tr}}
\newcommand{\slsh}[1]{/ \!\!\!\! #1}
\def\vac{|0\rangle}
\def\lvac{\langle 0|}
\def\hlf{\frac{1}{2}}
\def\ove#1{\frac{1}{#1}}
\def\Box{\square}
\def\ZZ{\mathbb{Z}}
\def\CC#1{({\bf #1})}
\def\bcomment#1{}
\def\bfhat#1{{\bf \hat{#1}}}
\def\VEV#1{\left\langle #1\right\rangle}
\newcommand{\ex}[1]{{\rm e}^{#1}} \def\ii{{\rm i}}
\def\rr{{\rm r}} \def\rs{{\rm s}}\def\rv{{\rm v}}
\def\ri{{\rm i}}\def\rj{{\rm j}}
\newcommand{\lrbrk}[1]{\left(#1\right)}
\newcommand{\sfrac}[2]{{\textstyle\frac{#1}{#2}}}

\def\Li{{\rm Li}_2}

\font\mybb=msbm10 at 12pt
\def\bb#1{\hbox{\mybb#1}}

\font\myBB=msbm10 at 18pt
\def\BB#1{\hbox{\myBB#1}}

\begin{document}

\begin{flushright}
Brown-HET-1592,
IPPP/10/30,
DCPT/10/60,
QMUL-PH-10--04
\end{flushright}

\vspace{3pt}

\begin{center}

{\Large \bf  A Surprise  in the Amplitude/Wilson Loop Duality }
\end{center}
\vspace{11pt}

\begin{center}
{\mbox {\bf Andreas Brandhuber$^{a}$, Paul Heslop$^{b}$, Panagiotis Katsaroumpas$^{a}$,}}  \\
\vspace{0.1cm} {\mbox {\bf \
Dung Nguyen$^{c}$, Bill Spence$^{a}$,  Marcus Spradlin$^{c}$  and Gabriele Travaglini$^{a}$}}%
\end{center}

\begin{center}
{\small \em
\begin{itemize}
\item[\ \ \ \ \ \ $^a$]
Centre for Research in String Theory\\
Department of Physics,
Queen Mary University of London\\
London, E1 4NS,
United Kingdom\\
\item[\ \ \ \ \ \ $^b$]
Institute for Particle Physics Phenomenology  \\
Department of Mathematical Sciences and Department of Physics\\
Durham University,
Durham,  DH1 3LE, United Kingdom
\item[\ \ \ \ \ \ $^c$]
Department of Physics\\
Brown University, Providence, Rhode Island 02912, USA

\end{itemize}
}

\vspace{-8pt}

\vspace{23pt} {\bf Abstract}
\end{center}

\noindent
One of the many remarkable features of MHV scattering amplitudes
is their conjectured equality
to lightlike polygon Wilson
loops, which apparently holds
at all orders in perturbation theory as well as non-perturbatively.
This duality is usually expressed in terms of purely
four-dimensional quantities obtained by appropriate
subtraction of the IR and UV divergences from amplitudes
and Wilson loops respectively.
In this paper we demonstrate, by explicit calculation, the completely
unanticipated fact that the equality continues to hold
at two loops
through
${\cal O}(\eps)$ in dimensional regularization for both the four-particle
amplitude
and the (parity-even part of the) five-particle amplitude.

\setcounter{page}{0}
\thispagestyle{empty}
\newpage


\section{Introduction }
\setcounter{footnote}{0}

Amongst the many remarkable features of the mathematical
structure of scattering amplitudes that have emerged
in the past several years, one of the most mysterious remains
the apparent equality between planar
maximally helicity violating (MHV)  scattering amplitudes and
lightlike Wilson loops in maximally supersymmetric Yang-Mills
(SYM) theory.  This new aspect of duality first emerged
in \cite{am}, where Alday and Maldacena argued using AdS/CFT
that the prescription for computing
scattering amplitudes at strong coupling was mechanically
identical to that for computing the expectation value
of a Wilson loop
over the closed contour obtained by gluing the momenta of the 
scattering particles back-to-back to form a polygon with lightlike edges.

Adhering to the principle that there is no such thing as a coincidence
in SYM theory, it was suggested in \cite{dks,bht} that
MHV amplitudes and Wilson loops might
be equal to each other not just at strong coupling, as the
work of Alday and Maldacena indicated, but perhaps
even order by order in perturbation theory.
This bold suggestion was confirmed by explicit calculations at
one loop for four particles in \cite{dks} and for any number of particles in \cite{bht},
and at two loops for four and five particles in \cite{dhks4,dhks5}.

Already for a few years prior to these developments
planar MHV amplitudes in SYM had come under close scrutiny
following the discovery of the ABDK relation \cite{Anastasiou:2003kj}, which expresses
the four-point two-loop amplitude as a certain quadratic polynomial
in the corresponding one-loop amplitude, a relation which was
later checked to hold also for the five-point two-loop amplitude \cite{Cachazo:2006tj,2l5pt}.
The all-loop generalization of the ABDK relation, known as the BDS
ansatz after the authors of \cite{bds},
expresses an appropriately defined infrared finite part of
the all-loop amplitude in terms of the exponential 
of the one-loop amplitude.  This proposal has also been completely
verified for the  three-loop four-point amplitude \cite{bds}, and partially
explored for the three-loop five-point amplitude \cite{svw}.

However it was shown in \cite{am2}  that the ABDK/BDS ansatz is incompatible
with strong coupling results in the limit of a very large number of particles,
and indeed it was found in \cite{seven}  that starting from six
particles and two loops
the ansatz is incomplete and the amplitude is given by the ABDK/BDS expression
plus a nonzero `remainder function' (an analytic expression for which was
obtained in \cite{DelDuca1,DelDuca2,Zhang:2010tr}).  
The breakdown of the ABDK/BDS ansatz beginning at six particles
can be understood on the basis of dual conformal symmetry \cite{magic,dhks}, which completely
determines the form of the four- and five-particle amplitudes but allows for an
arbitrary function of conformal cross-ratios beginning at $n=6$ \cite{dhks5,dhks6}.
While dual conformal invariance of SYM scattering amplitudes remains a conjecture
beyond one loop, it is necessary if the equality between amplitudes
and Wilson loops is to hold in general since
the symmetry translates to the manifest ordinary conformal invariance
of the corresponding Wilson loops.

Of course dual conformal symmetry alone does not imply the amplitude/Wilson loop
equality since they could differ by an arbitrary function of cross-ratios,
but miraculously precise agreement was found in \cite{seven,dhks6}  between the two
sides for $n=6$ particles at two loops.
Evidently some magical aspect of SYM theory is at work beyond the already remarkable
dual conformal symmetry.

This series of developments has opened up a number of interesting directions
for further work.  In this paper we turn our attention to a question which might have
seemed unlikely to yield an interesting answer:  does the amplitude/Wilson loop
equality hold beyond ${\cal O}(\eps^0)$ in the dimensional regularization
parameter $\eps$?
This question is motivated largely by the observation \cite{bht} that at one loop,
the four-particle amplitude is actually equal to the lightlike four-edged Wilson loop
to all orders in $\eps$ after absorbing an $\eps$-dependent  normalization factor. 
Furthermore,  the parity-even part of the five-particle amplitude is equal to the
corresponding Wilson loop  to all orders in $\eps$, again after absorbing 
the same normalization factor.%
\footnote{For  $n>5$ the Wilson loop calculation reproduces only the all orders in $\eps$
two-mass easy box functions, while the  corresponding $n$-point amplitude 
contains additional parity-odd as well as parity-even terms which vanish as $\eps \to 0$.  }
To our pleasant surprise we find a  positive answer to this question at two loops:
agreement between the $n=4$ and the parity-even part of the $n=5$ amplitude and the
corresponding Wilson loop continues to hold at ${\cal O}(\eps)$ up to an additive
constant which can be absorbed into various structure functions.

Let us emphasize that this is a rather striking result which cannot reasonably
be called a coincidence: at this order in $\eps$ the amplitudes and Wilson loops
we compute depend on all of the kinematic variables
in a highly nontrivial way, involving polylogarithmic
functions of degree 5.
It would be very interesting to continue exploring this miraculous agreement
and to understand the reason behind it.
Dual conformal invariance cannot help in this regard since the symmetry
is explicitly broken in dimensional regularization so it
cannot say anything about terms of higher order in $\epsilon$, but of course
as mentioned above already at ${\cal O}(\eps^0)$ there must be some mechanism
beyond dual conformal invariance at work.

The rest of the paper is organized as follows. 
In Section 2 we  review key aspects of one- and two-loop amplitudes and Wilson loops, in particular the ABDK/BDS ansatz and the correspondence between MHV amplitudes and Wilson loops. We also summarize our main results on the equality of amplitudes and Wilson loops  up to $\cO (\eps)$ at four and five points in \eqref{diff4ofremainders}, \eqref{diff5ofremainders}, respectively. 
In Section 3 we present the four- and five-point amplitudes at one and two loops, and in Section 4 the corresponding Wilson loops. Section 5 is devoted   to  discussing the numerical methods that  have been employed in order to perform our analysis. 
Finally, in Section 6 we  compare amplitude and Wilson loops, showing the agreement between these two quantities up to and including $\cO(\eps)$ terms.
Two appendices complete the paper. In the first one, we present results valid to all orders in $\eps$ for all Wilson loop diagrams contributing to the four-point case, with the exception of the so-called ``hard" diagram topology, which is evaluated only up to and including  $\cO(\eps)$ terms. In the second appendix, we present a novel expression for the all-orders in $\eps$ one-loop $n$-point Wilson loop diagrams which have simple analytic continuation properties.

\section{Review and Summary of Main Results }

The infinite sequence of $n$-point planar
maximally helicity violating
(MHV) amplitudes in $\cN\!=\!4$
super-Yang-Mills theory
(SYM) has a  remarkably simple structure.
Due to supersymmetric Ward identities 
\cite{Grisaru:1976vm,  Grisaru:1977px,Mangano:1990by,Dixon:1996wi}, 
at any loop order $L$, the amplitude can be expressed as the tree-level amplitude, times a scalar, helicity-blind function $\cM_n^{(L)}$:
\beq\label{fullampl}
\cA_{n}^{(L)} \ =  \cA^{\rm tree}_n\, \cM_n^{(L)}.
\eeq
In \cite{Anastasiou:2003kj}, ABDK discovered an intriguing  iterative structure
in the two-loop expansion of the MHV amplitudes at four points. This relation can be written as
\beq
\label{babis}
 \cM_4^{(2)}(\e)  - \frac{1}{2} \big( \cM_4^{(1)}(\e) \big)^2 \ = \ f^{(2)} (\e) \cM_4^{(1)} (2 \e ) + C^{(2)} + \cO (\e)
 \ ,
 \eeq
where IR divergences are regulated by working in $D = 4 - 2\e$
dimensions
(with $\e<0$),
 \beq
 \label{f2}
 f^{(2)} (\e) \ = \ -\zeta_2 - \zeta_3 \e - \zeta_4 \e^2
 \ ,
 \eeq 
 and
 \beq
 \label{cc2}
 C^{(2)}\  = \  -\frac{1}{2} \zeta_2^2
 \ .
 \eeq
The ABDK relation \eqref{babis} is built upon the known exponentiation
of infrared divergences \cite{ir7,ir8}, which guarantees that the
singular terms must agree on both sides of \eqref{babis}, as well
as on the known behavior of amplitudes under collinear limits
\cite{ku,vittorio}.  
The (highly nontrivial) content of the ABDK relation
is that \eqref{babis} holds exactly as written at ${\cal O}(\eps^0)$.
However, ABDK observed that the $\cO(\eps)$ terms do not satisfy the
same iteration relation \cite{Anastasiou:2003kj}.

In \cite{Anastasiou:2003kj}, it was  further conjectured  that \eqref{babis}
should hold for two-loop amplitudes with an arbitrary number of legs,
with the same quantities \eqref{f2} and \eqref{cc2} for any $n$.
In the five-point case, this conjecture was confirmed first in \cite{Cachazo:2006tj} 
for the parity-even part of the two-loop amplitude, and later 
in \cite{2l5pt} for the complete amplitude. Notice that for the iteration to be satisfied parity-odd terms that enter on
the left-hand side of the relation must cancel up to and including $\cO(\eps^0)$ terms, since the right-hand side is
parity even up this order in $\eps$. So far this has been checked and confirmed at two-loop order for five and six
particles \cite{2l5pt,Cachazo:2008hp}. This is also crucial for the duality with Wilson loops (discussed below) which by construction cannot produce 
parity-odd terms at two loops.

It has been found that starting from six particles and two loops,  the ABDK/BDS ansatz \eqref{babis} needs to be modified by allowing the presence of a remainder function $\cR_n$ \cite{seven,dhks6},
\beq
\label{fm2}
\cM^{(2)}_n(\e ) - \frac{1}{2} \Big( \cM^{(1)}_n (\e) \Big)^2 \ = \
f^{(2)} (\e) \cM^{(1)}_n  ( 2 \e ) \, + \,  C^{(2)} \, + \, \cR_n \
\, + \, E_n ( \eps ) 
\ ,
\eeq
where $\mathcal{R}_n$ is $\e$-independent
and $E_n$ vanishes as $\e \to 0$.  We parameterize the latter
as
\beq
\label{eenn}
E_n(\eps) = \eps \, \mathcal{E}_n +
\cO(\eps^2)\ . 
\eeq
 In this paper we will discuss in detail $\mathcal{E}_n$
for $n=4,5$ where we find a remarkable relation to the same quantity
calculated from the Wilson loop. Hitherto this relation was only expected
to hold for the finite parts of the remainder $\mathcal{R}_n$.

In a parallel development, Alday and Maldacena addressed the problem of calculating scattering amplitudes at strong coupling in $\cN=4$ SYM using the AdS/CFT correspondence. 
Their remarkable result showed that the planar amplitude at strong coupling is calculated by a Wilson loop 
\beq
\label{wil}
W[ \cC_n]  \ := \ {\rm Tr} \, \cP \exp \left[ i g\oint_{\cC_n} \! d\tau  \ \dot{x}^{\mu} (\tau )A_\mu (x(\tau ))   \right]
\ , 
\eeq
whose contour ${\cC_n}$  is the  $n$-edged polygon obtained by joining the lightlike momenta of the particles following the order induced by the colour structure of the planar amplitude. At strong coupling the calculation amounts to finding the minimal area of a surface ending on the contour ${\cC_n}$ embedded at the boundary of a T-dual $AdS_5$ space \cite{am}.
Shortly after, it was realised that the very same Wilson loop evaluated at weak coupling reproduces all one-loop MHV amplitudes in $\cN=4$ SYM  \cite{dks,bht}. The conjectured relation between MHV amplitudes and Wilson loops found
further strong support by explicit two loop calculations at four \cite{dhks4},   five \cite{dhks5} and six points \cite{dhks6, seven,Cachazo:2008hp}.
In particular, the absence of a non-trivial remainder function in the four- and five-point case was later explained in \cite{dhks5} from the Wilson loop perspective, where it was realised that the BDS ansatz is a solution to the anomalous Ward identity for the Wilson loop associated to the dual conformal symmetry \cite{magic}.

The Wilson loop in \eqref{wil} can be expanded in powers of the 't Hooft coupling $a\  := \  {g^2 N / (8 \pi^2)}$
as%
\footnote{We follow the definitions and conventions of \cite{Anastasiou:2009kn}, 
to which we refer the reader for more details.}
\beq
\label{nae}
\lan W[\cC_n ] \ran \  :=  1 \, + \, \sum_{l=1}^{\infty} a^l W^{(l)}_n \ := \   \exp \sum_{l=1}^{\infty} a^l w^{(l)}_n
\ .
\eeq
Note that the exponentiated form of the Wilson loop is guaranteed by the non-Abelian exponentiation theorem \cite{gatheral,taylor}. The $w^{(l)}_n$ are obtained from ``maximally non-Abelian" subsets of Feynman diagrams contributing to the $W^{(l)}_n$ and in particular from \eqref{nae} we find
\beq
w^{(1)}_n \ = \  W^{(1)}_n \, , \qquad
w^{(2)}_n \ = \  W^{(2)}_n \, - \, \frac{1}{2} \, (W^{(1)}_n)^2
\ .
\eeq
The UV divergences of the $n$-gon Wilson loop are
regulated by working in $D = 4 + 2 \epsilon$ dimensions with $\epsilon<0$.
The one-loop Wilson loop $w^{(1)}_n$ times the tree-level MHV amplitude is equal to the one-loop MHV amplitude, first calculated in \cite{bddk} using the unitarity-based approach \cite{fusing}, up to a regularization-dependent factor. 
This implies that non-trivial remainder functions can only appear at two and higher loops. At two loops, which is the
main focus of this paper,
we define the remainder function $\cR_n^{\rm WL}$ for an $n$-sided Wilson loop  as%
\footnote{We expect a remainder function at every loop order $l$ and the corresponding equations would be $w_n^{(l)}(\e)  = f_\mathrm{WL}^{(l)}(\epsilon) \, w_n^{(1)}(l \epsilon) \, + \, C_\mathrm{WL}^{(l)} \, +\, \cR_{n,\mathrm{WL}}^{(l)} \, + \, E_{n,\mathrm{WL}}^{(l)}(\eps)$.}
\begin{align}
\label{wbds}
  w_n^{(2)}(\e)  &= f_\mathrm{WL}^{(2)}(\epsilon) \, w_n^{(1)}(2 \epsilon) \, + \, C_\mathrm{WL}^{(2)} \, +\, \cR_n^\mathrm{WL} \, +\, 
E_n^{\mathrm{WL}}(\eps)
  \  ,
\end{align}
where
\beq \label{flepsW}
f_\mathrm{WL}^{(2)}(\epsilon ) :=\, f_0^{(2)} + f_{1,\mathrm{WL}}^{(2)} \epsilon + f_{2,\mathrm{WL}}^{(2)} \epsilon^2
\ .
\eeq
Note that $f_0^{(2)} =-  \zeta_2$, which  is the same as on the amplitude side, while $f_{1,\mathrm{WL}}^{(2)}= G_{\rm eik}^{(2)} = 7 \zeta_3 $  \cite{kk}.
In \cite{Anastasiou:2009kn}, the four- and five-edged Wilson loops were cast in the form \eqref{wbds}
and by making the natural requirements
\beq
\label{R45}
\cR_4^\mathrm{WL}\, = \, \cR_5^\mathrm{WL} \, = \, 0
\ ,
\eeq
this allowed for a determination of the coefficients $f_{2,\mathrm{WL}}^{(2)}$ and $C_\mathrm{WL}^{(2)}$.
The results found in\cite{Anastasiou:2009kn}, are%
\footnote{The $\cO (1)$ and $\cO (\e )$ coefficients of $f_\mathrm{WL}^{(2)}(\epsilon)$ had been determined earlier in
\cite{dhks4}.}
\beq
f_\mathrm{WL}^{(2)}(\epsilon)\ =\  -\zeta_2 \, + \,   7 \zeta_3 \, \e  \ - \    5\zeta_4\, \e^2
\ ,
\eeq
and
\beq
C_\mathrm{WL}^{(2)} \ =\  -\frac{1}{2} \zeta_2^2
\ .
\label{cc2WL}
\eeq
As noticed in  \cite{Anastasiou:2009kn}, there is an intriguing agreement between
the constant $C_\mathrm{WL}^{(2)}$ and the corresponding value of the same quantity on the amplitude side.

What has been observed so far is a duality between Wilson loops and
amplitudes up to finite terms. In turn this can be reinterpreted as an
equality of the corresponding remainder functions
\footnote{An
  alternative interpretation of the duality in terms of certain ratios
  of amplitudes (Wilson loops) has been given recently
  in~\cite{Heslop:2010xa}.}
\begin{align}
\label{dualitywa}
 \cR_n \ = \   \cR_n^\mathrm{WL}\ .
\end{align}
A consequence of the precise  determination of the  constants  $ f_{2,\mathrm{WL}}^{(2)}$ and
$C^{(2)}_\mathrm{WL}$ is that  no additional constant term is allowed on the right hand side of
\eqref{dualitywa}.  For the same reason,  the Wilson loop remainder function must then have
the same collinear limits as its amplitude counterpart,  i.e.
\beq
\label{abb}
  \cR_n^\mathrm{WL} \rightarrow \cR_{n-1}^\mathrm{WL} \ ,
\eeq
with no extra constant  on the right hand side of \eqref{abb}.

The main result of the present paper is that for $n=4,5$ the relation
between amplitudes and Wilson loops continues to hold for terms of
order $\eps^1$. In particular we find
\beq
\label{diff4ofremainders}
\mathcal{E}_4^{(2)} = \mathcal{E}_{4,\mathrm{WL}}^{(2)} - 3 \zeta_5 \ ,
\eeq
\beq
\label{diff5ofremainders}
\mathcal{E}_5^{(2)} = \mathcal{E}_{5,\mathrm{WL}}^{(2)} - \frac{5}{2} \zeta_5 \ .
\eeq
Note that these results have been obtained (semi-)numerically with typical
errors of   $10^{-8}$ at $n=4$  and $10^{-4}$ for $n=5$. 
Details of the
calculations are presented in the remaining sections of this paper.
More precisely $\mathcal{E}_4^{(2)}$ is known analytically \cite{bds},
while the analytic evaluation of $\mathcal{E}_{4,\mathrm{WL}}^{(2)}$
is discussed in appendix A. 
At five points all results are numerical and furthermore on the amplitude
side we only considered the parity-even terms. It is an interesting open
question whether the parity-odd terms cancel at $\cO(\eps)$ as they
do at $\cO(\eps^0)$ \cite{2l5pt}.

\section{Amplitudes}

In this section we review the ingredients necessary for our calculation
of the ${\cal O}(\eps)$ terms in the ABDK relation for the $n=4,5$ point
amplitudes.

\subsection{One-Loop Amplitudes }

We begin with the one-loop amplitudes, for which analytic results
can be given to all orders in $\epsilon$.

Following the conventions of~\cite{Anastasiou:2003kj}, the one-loop
four-point
amplitude may be expressed as~\cite{Green:1982sw}
\begin{equation}\label{m14}
{\cal M}^{(1)}_4 = - \frac{1}{2} s t I^{(1)}_4
\end{equation}
where $s=(p_1+p_2)^2$, $t=(p_2+p_3)^2$ are the usual
Mandelstam variables
and $I^{(1)}_4$ is the massless scalar box integral
\begin{equation}
I^{(1)}_4 =
{\hbox{\lower 40.pt\hbox{
\includegraphics{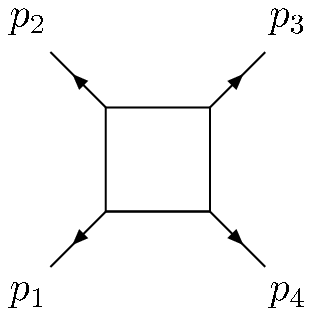}
}}}
= \frac{e^{\epsilon \gamma}}{i \pi^{D/2}} \int d^D p
\frac{1}{p^2 (p-p_1)^2 (p-p_1-p_2)^2 (p+p_4)^2},
\end{equation}
which we have written out in order to emphasize the
normalization convention (followed throughout this section) that
each loop momentum integral carries an overall factor
of $e^{\epsilon \gamma}/i \pi^{D/2}$.
The integral may be evaluated explicitly
(see for example~\cite{Bern:1993kr}) in terms of the ordinary
hypergeometric function ${}_2{\rm F}_1$,
leading to the exact expression
\begin{equation}
\label{eq:amplitudeone}
{\cal M}^{(1)}_4 = - 
\frac{e^{\epsilon \gamma}}{\epsilon^2}
\frac{\Gamma(1 + \epsilon) \Gamma^2(1 - \epsilon)}{\Gamma(1 - 2 \epsilon)}
\left[(-s)^{-\epsilon}
{}_2{\rm F}_1(1,-\epsilon,1-\epsilon,1+s/t) +
(s \leftrightarrow t) \right],
\end{equation}
valid to all orders in $\epsilon$.
We will always be studying the amplitude/Wilson loop duality in the
fully Euclidean regime where all momentum invariants such as $s$ and $t$
are negative.
The formula~(\ref{eq:amplitudeone}) applies in this regime
as  long as we are careful to navigate branch cuts
according to the rule
\begin{equation}
\label{eq:branches}
(-z)^{-\epsilon}\, {}_2{\rm F}_1(-\epsilon,-\epsilon,1-\epsilon,
1+z) := \lim_{\varepsilon \to 0} {\rm Re} \left[
\frac{{}_2{\rm F}_1(-\epsilon,-\epsilon,1-\epsilon,
1+z + i \varepsilon)}{(-z + i \varepsilon)^\epsilon} \right]
\end{equation}
when $z > 0$.

Five-point loop amplitudes ${\cal M}_5^{(L)}$ contain both parity-even
and parity-odd contributions
after dividing by the tree amplitude as in \eqref{fullampl}.
The parity-even part of the one-loop five-point
amplitude is given by~\cite{BDDKSelfDual}
\begin{equation}\label{m15}
{\cal M}_{5+}^{(1)} = - \frac{1}{4} \sum_{\rm cyclic} 
s_{3} s_{4} I^{(1)}_5, \qquad
I^{(1)}_5 =
{\hbox{\lower 50.pt\hbox{
\includegraphics{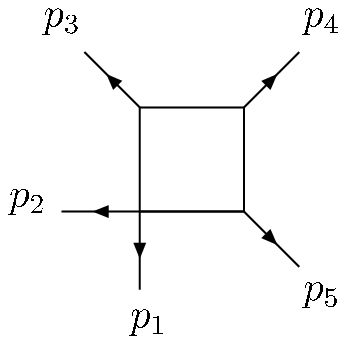}
}}},
\end{equation}
where $s_i = (p_i + p_{i+1})^2$ and the sum runs over the five
cyclic permutations of the external momenta $p_i$.
This integral can also be
explicitly evaluated (see for example~\cite{Bern:1993kr}),
leading to the all-orders in $\epsilon$ result
\begin{equation}
\begin{aligned}
{\cal M}^{(1)}_{5+} = - \frac{e^{\epsilon \gamma}}{\epsilon^2}
\frac{ \Gamma(1+\epsilon) \Gamma^2(1-\epsilon)}{\Gamma(1 - 2 \epsilon)}
\frac{1}{2} \sum_{\rm cyclic}
\Big[
\left(-\frac{s_1-s_4}{s_3 s_4}\right)^\epsilon
{}_2{\rm F}_1(-\epsilon,-\epsilon,1-\epsilon,
1 - \frac{s_3}{s_1 - s_4})\\
+\left(-\frac{s_1-s_3}{s_3 s_4}\right)^\epsilon
{}_2{\rm F}_1(-\epsilon,-\epsilon,1-\epsilon,
1 - \frac{s_4}{s_1 - s_3})\\
-\left(-\frac{(s_1-s_3)(s_1-s_4)}{s_1s_3s_4}\right)^\epsilon
{}_2{\rm F}_1(-\epsilon,-\epsilon,1-\epsilon,
1 - \frac{s_3 s_4}{(s_1-s_3)(s_1-s_4)})
\Big],
\end{aligned}
\label{eq:m1loop5}
\end{equation}
again keeping in mind \eqref{eq:branches}.

\subsection{Two-Loop Amplitudes}

\begin{figure}[t]
\begin{center}
\includegraphics{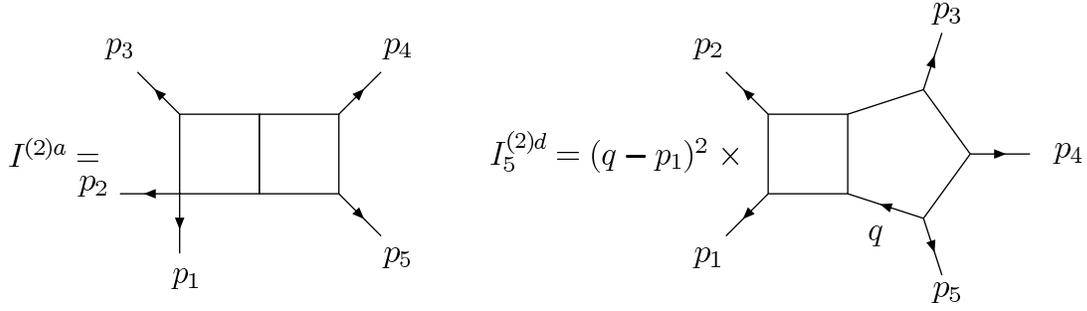}
\end{center}
\caption{\it Integrals appearing in the amplitude ${\cal M}^{(2)}_{5+}$. Note
that $I_5^{(2)d}$ contains the indicated
scalar numerator factor involving $q$, one of
the loop momenta.}
\label{2loop5fig}
\end{figure}

The two-loop four-point amplitude is expressed as~\cite{Bern:1997nh}
\begin{equation}
{\cal M}^{(2)}_4 =
\frac{1}{4} s^2 t
I^{(2)}_4 + (s \leftrightarrow t),
\qquad
I^{(2)}_4 =
{\hbox{\lower 40.pt\hbox{
\includegraphics{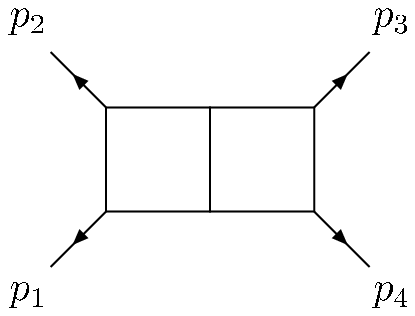}
}}},
\end{equation}
which may be evaluated analytically through ${\cal O}(\eps^2)$
using results from~\cite{bds} (no all-orders in $\eps$ expression
for the double box integral is known), from which we find
\begin{equation}
\begin{aligned}
\label{fourpointremainderamplitude}
\mathcal{E}_4 &=
 5\,{\rm Li}_5(-x) - 4 L\,{\rm Li}_4(-x)
+ \frac{1}{2} (3 L^2 + \pi^2) \, {\rm Li}_3(-x)
- \frac{L}{3} (L^2 + \pi^2) \, {\rm Li}_2(-x)
\\
& \qquad
- \frac{1}{24} (L^2 + \pi^2)^2 \log(1+x)
+ \frac{2}{45} \pi^4 L
- \frac{39}{2} \zeta_5
+ \frac{23}{12} \pi^2 \zeta_3,
\end{aligned}
\end{equation}
where $x = t/s$ and $L = \log x $.
A comment is in order here: In order to be able to present the amplitude
remainder \eqref{fourpointremainderamplitude}
in this form, we have pulled out a factor of $(s t )^{-L \eps/2}$
from each loop amplitude ${\cal M}^{(L)}_4$.  This renders
the amplitudes, and hence the ABDK remainder $E_4(\e)$,  dimensionless
functions of the single variable $x$.
We perform this step in the four-point case only,
where we are able to present analytic results for the amplitude 
and Wilson loop remainders.

The parity-even part of the two-loop five-point amplitude involves
the two integrals shown in  Figure~\ref{2loop5fig}, in terms of
which~\cite{Bern:1997it,Cachazo:2006tj,2l5pt}
\begin{equation}
{\cal M}^{(2)}_{5+} =
 \frac{1}{8} \sum_{\rm cyclic}
\left( s_3 s_4^2 I^{(2)a} + (p_i \to p_{6-i})\right)
+ s_1 s_3 s_4 I^{(2)d},
\end{equation}
where $s_i = (p_i + p_{i+1})^2$.
To evaluate this amplitude to ${\cal O}(\eps)$ we must resort
to a numerical calculation using
Mellin-Barnes parameterizations of the integrals
(which may be found for example in~\cite{Cachazo:2006tj}), which
we then expand through $\cO(\epsilon)$, simplify, and numerically
integrate with the help of the {\tt MB}, {\tt MBresolve},
and {\tt barnesroutines}
programs~\cite{Czakon:2005rk,Smirnov:2009up},
In this manner we have determined
the ${\cal O}(\eps)$ contribution ${\mathcal{E}}_5^{(2)}$
to the
five-point ABDK relation 
numerically
at a variety of kinematic points.  The results are displayed in 
Table \ref{tbl:remainders}. 

\section{Wilson Loops}

\subsection{One-Loop Wilson Loops} 

The one-loop Wilson loop was found in \cite{bht} for any number of
edges and to all orders in the dimensional regularization parameter $\epsilon$. 
It is obtained by summing over diagrams with a single gluon propagator stretching between any two edges of the
Wilson loop polygon. 
\begin{figure}[h!]
  \centering
  \begin{tabular}{ccc}
    \scalebox{0.3}{ \fcolorbox{white}{white}{
        \begin{picture}(204,194) (541,-151)
          \SetWidth{0.5}
          \SetColor{Black}
          \Gluon(595,-74)(672,-14){7.5}{5.54}
          \Gluon(595,-74)(672,-14){7.5}{5.54}
          \ArrowLine(541,43)(627,-151)
          \ArrowLine(745,-44)(542,43)
        \end{picture}
      }
    }&$\qquad \qquad$
    &
    \scalebox{0.3}
    {
      \fcolorbox{white}{white}{
        \begin{picture}(204,185) (132,-156)
          \SetWidth{0.5}
          \SetColor{Black}
          \ArrowLine(132,27)(329,27)
          \ArrowLine(137,-155)(336,-156)
          \Gluon(227,-155)(228,26){7.5}{11.5}
        \end{picture}
      }
    }
    \\
    cusp :=${\Gamma ( 1 + \e) e^{  \epsilon \gamma
   }}  \times \left( - {1\over 2 \epsilon^2} (-s_i )^{-\epsilon}\right)$
    &&
    finite :=${\Gamma ( 1 + \e) e^{  \epsilon \gamma
   }}  \times {\cal F}_\epsilon$
  \end{tabular}
\caption{\it One-loop Wilson loop diagrams. The expression of $\mathcal{F}_\eps$ is given in \eqref{fin1} of equivalently in \eqref{fin2}. 
}
\end{figure}
Diagrams with the propagator stretching between
adjacent edges $p_i$ and $p_{i+1}$ are known as cusp diagrams, and give the
infrared-divergent terms in the Wilson loop, proportional to
$(-2p_i\cdot p_{i+1})^{-\epsilon}/\epsilon^2=(-s_i)^{-\epsilon}/\epsilon^2$.

On the other hand,
diagrams for which the propagator stretches between two non-adjacent
edges are finite. Their contribution to the Wilson loop can be found
to all orders in $\eps$  and is (up to an $\eps$-dependent
factor) precisely equal to the finite part of a two-mass
easy or one-mass box function \cite{bht} (for details see appendix~\ref{appb}).
The general $n$-point one loop amplitude is given by the
sum over precisely these two-mass easy and one-mass box
functions~\cite{bddk} to $\cO(\epsilon^0)$.%
\footnote{The all-orders in
  $\epsilon$ $n$-point
  amplitude contains new integrals contributing at $\cO(\epsilon)$.} 
Thus we conclude that the Wilson
loop is equal to the amplitude at one loop for any $n$ up to finite
order in $\epsilon$ only (and up to a kinematic independent factor).

However at four and five points a much stronger statement can be
made. The  four-point amplitude and the parity-even part of the
five-point amplitude are both given by the sum over zero- and one-mass boxes 
 {\em   to all orders in $\epsilon$}. Thus the Wilson loop correctly reproduces
these one-loop amplitudes to all orders in $\epsilon$.  Using the
results in appendix B we find 
that the four-point Wilson loop (in a form which is manifestly real in
the Euclidean regime $s,t<0$) is given by 
\begin{align}
  W_4^{(1)}=&\ {\Gamma ( 1 + \e) e^{  \epsilon \gamma
   }} \Big\{ -{1\over \epsilon^2} \left[ (-s)^{-\epsilon}
  +(-t)^{-\epsilon}\right]  +  {\cal F}_\epsilon(s,t,0,0)+ {\cal
  F}_\epsilon(t,s,0,0)\Big\} \nonumber\\[10pt]
=&\ {\Gamma ( 1 + \e) e^{  \epsilon \gamma
   }} \Bigg \{ -{1\over \epsilon^2} \left[ (-s)^{-\epsilon}
  +(-t)^{-\epsilon}\right]  \nonumber \\
+&{1 \over \epsilon^2} \left({u\over s t}\right)^{\epsilon}  \bigg[  \left({t\over
       s}\right)^{\epsilon}  {}_2
    {\rm F}_1(\epsilon,\epsilon;1+\epsilon;-t/s)+ \left({s\over
       t}\right)^{\epsilon} {}_2
    {\rm F}_1(\epsilon,\epsilon;1+\epsilon;-s/t)-
2\pi\epsilon\cot(\epsilon\pi)
\bigg]\Bigg \}
\ . 
\label{4pnt1loop}
\end{align}
Note in particular the additional cotangent term explained in detail
at the end of appendix B. 
The generic form of the function $\cF_\eps $ is given in \eqref{fin1} of equivalently in \eqref{fin2}.

For the five-point amplitude we display a new form which has a simple
analytic continuation in all kinematical regimes and also a very
simple expansion in terms of Nielsen polylogarithms
(see~\eqref{3f2ex}). It is given in terms of ${}_3 
{\rm F}_2$ hypergeometric functions and is derived in detail in appendix B:
  \begin{align}
 W_5^{(1)}=& \sum_{i=1}^5 {\Gamma ( 1 + \e) e^{  \epsilon \gamma
   }} \Big[ -{1\over 2\epsilon^2} (-s_{i})^{-\epsilon}
   +  {\cal F}_\epsilon(s_{i},s_{i+1},s_{i+3},0)\Big] \nonumber\\[10pt]
=&\sum_{i=1}^5{\Gamma ( 1 + \e) e^{  \epsilon \gamma
   }} \Bigg \{  -{1\over 2\epsilon^2}
 (-s_{i})^{-\epsilon}\nonumber \\
&\qquad \qquad  \qquad \quad -{\frac12 \left({s_{i+3}-s_i-s_{i+1} \over s_i s_{i+1} }\right)^{\epsilon}}
 \bigg[\tfrac{s_{i+3}-s_i}{s_{i+1}}\,  {}_3 
    {\rm F}_2\big(1,1,1+\epsilon;2,2;\tfrac{s_{i+3}-s_i}{s_{i+1}}\big) \nonumber \\
&\qquad \qquad \qquad \qquad \qquad \qquad \qquad \qquad +
   \tfrac{s_{i+3}-s_{i+1}}{s_{i}} \,{}_3
    {\rm F}_2\big(1,1,1+\epsilon;2,2;\tfrac{s_{i+3}-s_{i+1}}{s_{i}}\big)  \nonumber \\
 & 
\qquad \qquad+\frac{H_{-\epsilon}}{\epsilon} \  - \ \tfrac{(s_{i+3}-s_i)(s_{i+3}-s_{i+1})}{s_i s_{i+1}} \, {}_3
    {\rm F}_2\big(1,1,1+\epsilon;2,2;\tfrac{(s_{i+3}-s_i)(s_{i+3}-s_{i+1})}{s_i s_{i+1}}\big)
\bigg]\Bigg \}
\label{5pnt1loop}
\end{align}  
where $H_n$ is the $n^{\rm th}$-harmonic number. Using hypergeometric identities
one can show that (up to the prefactor) the four- and five-sided
Wilson loops \eqref{4pnt1loop}, \eqref{5pnt1loop} are equal to the
four-point and the (parity-even part of the) five-point amplitudes
of~(\ref{eq:amplitudeone}) and~(\ref{eq:m1loop5}). 

The precise relation between the Wilson loop and the amplitude is
\begin{equation}
{W}^{(1)}_4 
\ = \ {\Gamma(1-2\eps)\over \Gamma^2(1-\eps)}
{\cal M}^{(1)}_4 
\ , 
\qquad \qquad {W}^{(1)}_5 
\ =  \ {\Gamma(1-2\eps)\over \Gamma^2(1-\eps)}
{\cal M}^{(1)}_{5+}
\ , 
\end{equation}
where ${\cal M}^{(1)}_4$ is the one-loop four-point amplitude and
${\cal M}^{(1)}_{5+}$ is the parity-even part of the five-point
amplitude. 

\subsection{Two-Loop Wilson Loops}

At two-loop order, the $n$-point Wilson loop is given by a sum over
six different types of
diagrams. 
These are described in general for polygons with any number of edges
in \cite{Anastasiou:2009kn} and are displayed for illustration below.  

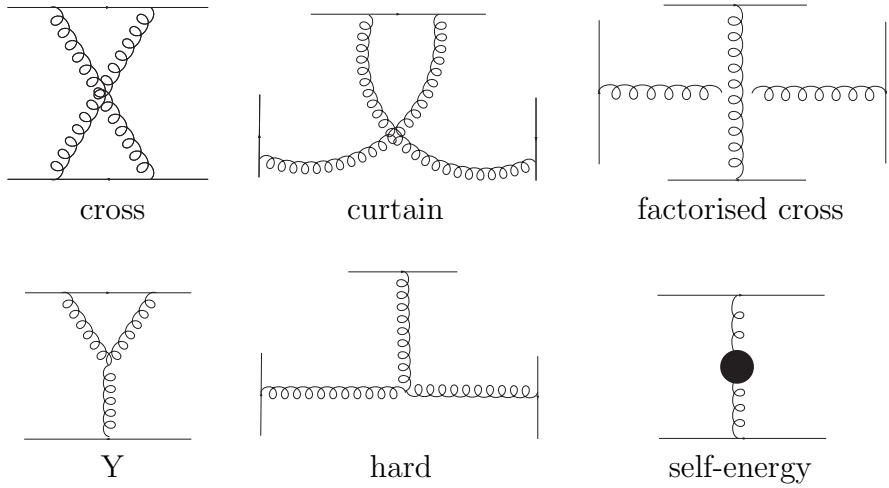
\begin{figure}[h!]
  \centering
  \begin{tabular}{ccc}
  \scalebox{0.3}{
\fcolorbox{white}{white}{
  \begin{picture}(252,221) (12,-10)
    \SetWidth{0.5}
    \SetColor{Black}
    \ArrowLine(12,208)(264,208)
    \ArrowLine(264,-8)(12,-8)
    \Gluon(66,208)(192,-8){9}{13.71}
    \Gluon(192,208)(66,-8){9}{13.71}
    \ArrowLine(12,208)(264,208)
    \ArrowLine(264,-8)(12,-8)
    \Gluon(66,208)(192,-8){9}{13.71}
    \Gluon(192,208)(66,-8){9}{13.71}
  \end{picture}
}
}&
  \scalebox{0.3}{
\fcolorbox{white}{white}{
  \begin{picture}(350,212) (187,-118)
    \SetWidth{0.5}
    \SetColor{Black}
    \ArrowLine(189,-113)(190,-9)
    \ArrowLine(537,-13)(537,-117)
    \ArrowLine(189,-113)(190,-9)
    \ArrowLine(537,-13)(537,-117)
    \GlueArc(470.3,42.33)(151.34,161.24,295.73){-7.5}{24.1}
    \GlueArc(249.98,65.93)(167.07,-111.04,8.98){-7.5}{23.47}
    \ArrowLine(254,92)(474,92)
  \end{picture}
}
}&
  \scalebox{0.3}{
\fcolorbox{white}{white}{
  \begin{picture}(363,223) (332,-11)
    \SetWidth{0.5}
    \SetColor{Black}
    \ArrowLine(594,-10)(420,-10)
    \ArrowLine(334,10)(334,191)
    \ArrowLine(694,189)(694,11)
    \ArrowLine(415,210)(595,210)
    \Gluon(505,210)(506,-8){9}{10.29}
    \Gluon(333,101)(488,100){9}{6.86}
    \Gluon(526,99)(695,99){9}{7.71}
  \end{picture}
}
}\\
$\qquad$ cross$\qquad$&$\qquad$curtain$\qquad$&$\qquad$factorised cross$\qquad$\\[.5cm]
  \scalebox{0.3}{
\fcolorbox{white}{white}{
  \begin{picture}(210,187) (38,-371)
    \SetWidth{0.5}
    \SetColor{Black}
    \Gluon(85,-187)(145,-278){7.5}{6.06}
    \Gluon(144,-278)(204,-187){7.5}{6.06}
    \Gluon(145,-368)(145,-278){7.5}{5}
    \ArrowLine(39,-187)(248,-187)
    \ArrowLine(248,-371)(38,-371)
  \end{picture}
}
}&
  \scalebox{0.3}{
\fcolorbox{white}{white}{
  \begin{picture}(350,211) (761,-14)
    \SetWidth{0.5}
    \SetColor{Black}
    \ArrowLine(873,195)(1010,195)
    \ArrowLine(763,-11)(764,93)
    \ArrowLine(1111,89)(1111,-15)
    \Gluon(941,196)(944,45){7.5}{9.36}
    \Gluon(763,43)(945,43){7.5}{11.57}
    \Gluon(1111,44)(946,46){7.5}{10.36}
  \end{picture}
}
}&
  \scalebox{0.3}{
\fcolorbox{white}{white}{
  \begin{picture}(212,183) (311,-379)
    \SetWidth{0.5}
    \SetColor{Black}
    \ArrowLine(311,-198)(521,-198)
    \ArrowLine(523,-378)(313,-378)
    \Gluon(413,-378)(413,-301){7.5}{3.49}
    \Gluon(412,-273)(412,-198){7.5}{2.57}
    \Vertex(411,-288){21.21}
  \end{picture}
}
}\\
$\qquad$ Y$\qquad$&$\qquad$ hard$\qquad$&$\qquad$self-energy$\qquad$
\end{tabular}
  \caption{\it The six different diagram topologies contributing to the
    two-loop Wilson loop. For details see \cite{Anastasiou:2009kn}.}
\label{fig:2loops}
\end{figure}

The computation of the four-point two-loop Wilson loop up to
$\cO(\eps^0) $ was first performed in \cite{dhks4}.  In appendix
\ref{4ptallorders} we display all the contributing diagrams for this
case
and give expressions for these to all
orders in $\epsilon$ in all cases  except for the ``hard" diagram,
which we  give up to and including  terms of $\cO( \epsilon) $. 
Summing up the contributions from all these diagrams we obtain the
result for the two-loop four-point Wilson loop to $\cO(\epsilon)$. This
is displayed in \eqref{wcc}  of the next subsection.

The five-point two-loop Wilson loop  was calculated up to  $\cO( \epsilon^0) $ 
in \cite{dhks5}. 
In order to obtain results at one order higher in $\epsilon$ we have
proceeded by using numerical methods. In particular we have used
Mellin-Barnes techniques to evaluate and expand all the two-loop integrals 
of Figure \ref{fig:2loops}. 
This is described in more detail in Section \ref{MBintegration}.

\subsubsection{The Complete Two-Loop Wilson Loop at Four Points}

Here is our final result for the four-point Wilson loop at two loops  expanded up to and including terms of $\cO(\eps)$: 
\beqa
\label{wcc}
w^{(2)}_4 &=& \cC \times
\Big[(-s)^{- 2\eps}+(-t)^{- 2\eps}\Big]
\times\Big[ {w_2\over \eps^2}+{w_1\over \eps} +w_0 +w_{-1}\eps+\cO(\eps^2)\Big]\ , 
\eeqa
where
\beqa
\label{w_2}
w_2&=&{\pi^2\over 48}\ , \\
w_1&=&- {7\zeta_3\over 8}\ , \\
w_0&=&-\frac{\pi^2}{48}  \left(\log ^2 x +\pi ^2\right) +{\pi^4\over
  144}\, = \,  -\frac{\pi^2}{48}  \left(\log^2 x +{2\over 3} \pi ^2\right)
  \ , \\ 
  \label{w-1}
w_{-1} &= &  -  {1\over 1440} \Big[
-46 \pi^4 \log x  - 10 \pi^2 \log^3 x + 
  75 \pi^4 \log (1 + x)  + 90 \pi^2 \log^2 x  \log (1 + x) 
  \nonumber\\
&&
  +15 \log^4 x \log( 1 + x) + 240 \pi^2 \log x  \, {\rm Li}_2 (-x)  + 
  120 \log^3 x \, {\rm Li}_2 (-x)  
  \nonumber\\
&& - 300 \pi^2  {\rm Li}_3 (-x) - 
  540 \log^2 x \,  {\rm Li}_3 (-x) + 1440 \log x \, {\rm Li}_4 (-x)
  \nonumber\\
&&
 -  1800 {\rm Li}_5 (-x) - 1560\pi^2  \zeta_3  - 
  1260 \log^2 x  \, \zeta_3  + 5940 \zeta_5 
  \Big]\ , 
\eeqa 
and 
\beq
\label{multby}
\cC:=2 
\Big[{\Gamma ( 1 + \e) e^{  \gamma \e}}\Big]^2
\ = \ 2 
\Big ( 1 + \zeta_2 \e^2 - {2\over 3} \zeta_3 \e^3 \Big) \, + \, \cO (\e^4 )
\ . 
\eeq
We recall that $x = t/s$. 

We would like to point out the simplicity of our result \eqref{wcc} -- 
specifically,  \eqref{w_2}--\eqref{w-1} are expressed only in terms of standard polylogarithms. Harmonic polylogarithms and Nielsen polylogarithms are present in the expressions of separate Wilson loop diagrams, as can be seen in appendix A, but cancel after 
summing all contributions. 

\subsubsection{The $\cO(\e)$ Wilson Loop  Remainder Function at Four Points}

Using the result \eqref{wcc} and the one-loop expression for the Wilson loop, one can work out the expression 
for the remainder function at $\cO (\eps)$, as defined in \eqref{fm2} and \eqref{eenn}.  
Our result is
\beqa
\label{rem4ptwl}
 \cE_{4, \rm WL}  &=&
{ 1 \over 360}  \, \bigg[ 
 16 \pi^4 \log x  - 15 \pi^4 \log (1 + x) - 
   30 \pi^2 \log^2 x \log (1 + x)  \nonumber \\
   && - 15 \log^4 x  \log ( 1 + x )  - 
   120 \pi^2 \log  x  {\rm Li}_2 ( -x)  - 120 \log^3  (x)  {\rm Li}_2 ( -x) 
   \nonumber \\
   &&+ 
   180 \pi^2 {\rm Li}_3 ( -x) + 540 \log^2 x  {\rm Li}_3( -x)  - 
   1440 \log (x)  {\rm Li}_4 ( -x)\nonumber  \\
   && + 1800 {\rm Li}_5 ( -x)  + 
   690 \pi^2 \zeta_3 - 5940 \zeta_5
   \bigg]
\ , 
\eeqa
where we recall that  $\cE_{n, \rm WL}$ is related to the quantity $E_n$ introduced in \eqref{fm2}  and   \eqref{eenn}. 
Remarkably, \eqref{rem4ptwl} does not contain any harmonic polylogarithms. We will compare the Wilson loop remainder 
\eqref{rem4ptwl} to the corresponding amplitude remainder \eqref{fourpointremainderamplitude}
in Section \ref{4ptremcomp}.%
\footnote{ Similarly to what was done for the amplitude remainder \eqref{fourpointremainderamplitude}, in arriving at 
\eqref{rem4ptwl} we have pulled out a factor of $(s t )^{-\eps/2}$  per loop in order to obtain a result which depends only on the ratio $x:= t/s$.}

\subsubsection{The $\cO(\e)$ Wilson Loop  at  Five Points and the Five-Point Remainder Function}

For the five-point amplitude and Wilson loop  at two loops we resort to completely numerical evaluation of the contributing integrals, and a comparison of the remainder functions is then performed. We postpone this discussion to 
section \ref{5ptsectioncomp}.

\section{Mellin-Barnes Integration}
\label{MBintegration}
   The two-loop five-point Wilson loop and amplitude have been numerically evaluated by
  means of the Mellin-Barnes (MB) method using the {\tt MB} package
  ~\cite{Czakon:2005rk} in {\tt MATHEMATICA}.
  At the heart of the method lies the Mellin-Barnes representation
  \begin{equation}\label{eq:MBrep2terms}
    \frac{1}
         {(X+Y)^{\lambda}}
    =
      \frac{1}
           {2 \pi i}
      \frac{1}
           {\Gamma (\lambda)}
      \int_{-i \infty}^{+i \infty}
      dz
      \:
        \frac{X^z}
             {Y^{\lambda+z}}
        \Gamma(-z)
        \Gamma(\lambda+z).
  \end{equation}
  We will use the integral representation for the hard diagram of the Wilson loop
  as an example in order to describe 
  the procedure we followed. The integral for the specific diagram  shown 
  in Figure \ref{generalharddiagram} has the expression 
    \begin{alignat}{1}
   \label{hd}
    &f_{\text{H}}(p_1,p_2,p_3;Q_1,Q_2,Q_3)
 \\
    &\phantom{=}
    =
    \frac{1}{8}
    \frac{\Gamma (2+2 \epsilon)}{\Gamma (1+ \epsilon)^2}
    \int_0^1(\prod_{i=1}^{3}{d\tau_i})
    \int_0^1(\prod_{i=1}^{3}{d\alpha_i})
    \delta (1-\sum_{i=1}^{3}{\alpha_i})
    (\alpha_1\alpha_2\alpha_3)^{\epsilon}
    \frac{\mathcal{N}}
         {\mathcal{D}^{2+2\epsilon}}.
     \nonumber
  \end{alignat}
  We write the numerator and denominator as a function of the 
  momentum invariants, i.e.~squares of sums of consecutive momenta,
  \begin{alignat}{1}
    \mathcal{D}
    &=
      -\alpha_1 \alpha_2 \left[ (p_1+Q_3+p_2)^2 (1 - \tau_1)\tau_2
      + (p_1+Q_3)^2 (1 - \tau_1) (1 - \tau_2)\right.
    \nonumber\\
    &\phantom{=}
     \qquad\qquad
      \left.
      +(Q_3+p_2)^2 \tau_1 \tau_2 
      + Q_3^2 \tau_1 (1 - \tau_2)
      \right]+\text{cyclic}(1,2,3),
  \end{alignat}
  \begin{alignat}{1}
    \mathcal{N}
    &=
     2 \left[2 (p_1 p_2) (p_3 Q_3)
              -(p_2 p_3) (p_1 Q_3)-(p_1 p_3) (p_2 Q_3)\right] 
     \alpha_1 \alpha_2
    \nonumber\\
    &\phantom{=}
     +2 (p_1 p_2) (p_3 p_1) 
     \left[\alpha_1 \alpha_2 (1 - \tau_1)+\alpha_3 \alpha_1 \tau_1\right]
     +\text{cyclic}(1,2,3),
  \end{alignat}
  where
  \begin{alignat}{1}
    2 p_i p_{i+1}
    &=
      -(p_i+Q_{i+2})^2
      +Q_{i+2}^2
      -(Q_{i+2}+p_{i+1})^2
      +(Q_i+p_{i+2}+Q_{i+1})^2,
  \nonumber   \\
    2 p_i Q_i
    &=
      -(p_i+Q_{i+2}+p_{i+1})^2
      +(Q_{i+2}+p_{i+1})^2
    \nonumber\\
    &\phantom{=}
      -(p_{i+2}+Q_{i+1}+p_i)^2
      +(p_{i+2}+Q_{i+1})^2,
\nonumber     \\
    2 p_i Q_j
    &=
      (p_i+Q_j)^2
      -Q_j^2.
  \end{alignat}
  By means of the substitution
  $\alpha_1 \rightarrow 1-\tau_4$,
  $\alpha_2 \rightarrow \tau_4 \tau_5$ and
  $\alpha_3 \rightarrow \tau_4 (1-\tau_5)$, we eliminate one integration and the delta 
  function to get a five-fold integral over $\tau_i\in[0,1]$. Next, we obtain
  an MB representation using the generalisation of (\ref{eq:MBrep2terms})
  \begin{equation}\label{eq:MBrepGeneral}
    \frac{1}
         {(\sum_{s=1}^m X_s)^{\lambda}}
    =
      \frac{1}
           {(2 \pi i)^{m-1}}
      \frac{1}
           {\Gamma (\lambda)}
      \left(
      \prod_{s=1}^{m-1}
      \int_{-i \infty}^{+i \infty}
      dz_s
      \right)
      \:
        \frac{\prod_{s=1}^{m-1} X_s^{z_s} \Gamma(-z_s)}
             {X_m^{\lambda+\sum_{s=1}^{m-1} z_s} \Gamma(\lambda+\sum_{s=1}^{m-1} z_s)},
  \end{equation}
  which introduces $m-1$ MB integration variables $z_s$, where $m$ is the number of
  terms in the denominator.
  At this point, the integrations over the $\tau_i$'s can be easily performed by means of
  the substitution
  \begin{equation}
    \int_0^1 dx \:
      x^{\alpha} (1-x)^{\beta}
    =
    \frac{\Gamma(\alpha+1)\Gamma(\beta+1)}
         {\Gamma(\alpha+\beta+2)}
         \, .
  \end{equation} 
  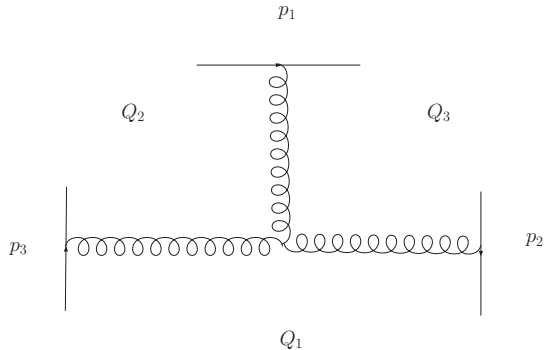
\begin{figure}[h]
\begin{center}
\scalebox{0.45}
{
\fcolorbox{white}{white}{
  \begin{picture}(464,293) (143,-64)
    \SetWidth{0.5}
    \SetColor{Black}
    \ArrowLine(299,174)(436,174)
    \ArrowLine(189,-32)(190,72)
    \ArrowLine(537,68)(537,-36)
    \Gluon(367,175)(370,24){7.5}{9.36}
    \Gluon(189,22)(371,22){7.5}{11.57}
    \Gluon(537,23)(372,25){7.5}{10.36}
    \Text(143,15)[lb]{\Large{\Black{$p_3$}}}
    \Text(369,213)[lb]{\Large{\Black{$p_1$}}}
    \Text(577,22)[lb]{\Large{\Black{$p_2$}}}
    \Text(370,-64)[lb]{\Large{\Black{$Q_1$}}}
    \Text(237,127)[lb]{\Large{\Black{$Q_2$}}}
    \Text(494,127)[lb]{\Large{\Black{$Q_3$}}}
  \end{picture}
}}
\end{center}
\caption{\it The hard diagram corresponding to \eqref{hd}.}
\label{generalharddiagram}
\end{figure}
  We are now left with an integrand that is an analytic function containing
  powers of the momentum invariants 
  $(-s_{ij})^{f(\{z_s\},\epsilon)}$ and Gamma functions 
  $\Gamma(g(\{z_i\},\epsilon))$, where $f$ and $g$ are
  linear combinations of the $z_s$'s and $\epsilon$.
  In order to perform the MB integrations, one has to pick appropriate
  contours, so that for each $z_s$ the $\Gamma(\cdots+z_s)$ poles are to the
  left of the contour and the $\Gamma(\cdots-z_s)$ poles are to the right.
  
  At this point we use various {\tt Mathematica} packages to perform 
  a series of operations in an automated way to finally obtain a 
  numerical expression at specific kinematic points. We will briefly summarise the 
  steps followed, while for more details we 
  refer the reader to the references documenting these packages and references
  therein.
  Using the {\tt MBresolve} package~\cite{Smirnov:2009up}, we pick appropriate
  contours and resolve
  the singularity structure of the integrand in \(\epsilon\). The latter
  involves taking residues and shifting contours, and is essential in
  order to be able to Laurent expand the integrand in \(\epsilon\).
  Using the {\tt barnesroutines} package 
\cite{Czakon:2005rk,Smirnov:2009up}, we 
  apply the Barnes lemmas, which in general generate more integrals
  but decrease their dimensionality, leading to higher precision results.
  Finally, using the {\tt MB} package \cite{Czakon:2005rk} we numerically integrate at specific Euclidean
  kinematic points to obtain a numerical expression. While all manipulations of the 
  integrals and the expansion in \(\epsilon\) are performed in {\tt Mathematica}, 
  the actual numerical integration for each term is  performed using the {\tt CUBA} 
  routines~\cite{Hahn:2004fe}
  for multidimensional numerical integration in {\tt FORTRAN}. The high number
  of diagrams, and number of integrals for each diagram, 
  makes the task of running the {\tt FORTRAN} integrations ideal for parallel computing.

\section{Comparison of the Remainder Functions}

\subsection{Four-point Amplitude and  Wilson Loop Remainders} 
\label{4ptremcomp}
The remainder functions for the four-point amplitude and Wilson loops are given in 
\eqref{fourpointremainderamplitude} and \eqref{rem4ptwl}, respectively. 
From these relations, it follows that the difference of remainders is a constant, $x$-independent term: 
\beqa
\cE_{4} \, = \, \,  \cE_{4, \rm WL}\, - \, 3 \,  \zeta_5\, 
\ , 
\label{remoeps}
\eeqa
as anticipated in  \eqref{diff4ofremainders}.

We would like to stress that this is a highly nontrivial result since there is no reason a priori to expect that the four-point remainder on the amplitude and Wilson loop side, \eqref{fourpointremainderamplitude} and \eqref{rem4ptwl} respectively,  agree (up to a constant shift). 
For example, anomalous dual conformal invariance is known to determine  the form of the four- and five-point Wilson loop only up to $\cO (\eps^0)$ terms \cite{dhks5}, but does not constrain terms which vanish as $\eps\to 0$.  
The expressions we derived for the amplitude and Wilson loop four-point remainders at $\cO (\eps)$ are also pleasingly simple, in that they only contain standard polylogarithms.

\subsection{Five-point Amplitude and Wilson Loop Remainders} 
\label{5ptsectioncomp}

  \begin{table}
  \scriptsize
    \noindent\center\(
    \begin{array}{|c|c|c|c|}
      \hline
      \# & (s_{12},s_{23},s_{34},s_{45},s_{51}) & \mathcal{E}_5^{(2)}
         & \mathcal{E}_{5,\mathrm{WL}}^{(2)}\\
      \hline
      1 & (-1,-1,-1,-1,-1) 
        & -8.463173 \pm 0.000047 & -5.8705280 \pm 0.0000068 \\
      2 & (-1,-1,-2,-1,-1) 
        & -8.2350 \pm 0.0024 & -5.64560 \pm 0.00063 \\
      3 & (-1,-2,-2,-1,-1) 
        & -7.7697 \pm 0.0026 & -5.17647 \pm 0.00076 \\
      4 & (-1,-2,-3,-4,-5) 
        & -6.234809 \pm 0.000032 & -3.642125 \pm 0.000018 \\
      5 & (-1,-1,-3,-1,-1) 
        & -8.2525 \pm 0.0027 & -5.65919 \pm 0.00097 \\
      6 & (-1,-2,-1,-2,-1) 
        & -8.142702 \pm 0.000023 & -5.5500050 \pm 0.0000092 \\
      7 & (-1,-3,-3,-1,-1) 
        & -7.6677 \pm 0.0034 & -5.0784 \pm 0.0013  \\
      8 & (-1,-2,-3,-2,-1) 
        & -6.8995 \pm 0.0029 & -4.31395 \pm 0.00093  \\
      9 & (-1,-3,-2,-5,-4) 
        & -6.9977 \pm 0.0031 & -4.40806 \pm 0.00099 \\
      10 & (-1,-3,-1,-3,-1) 
         & -8.2759 \pm 0.0025 & -5.69086 \pm 0.00085  \\
      11 & (-1,-4,-8,-16,-32) 
         & -8.7745 \pm 0.0078 & -6.1825 \pm 0.0051 \\
      12 & (-1,-8,-4,-32,-16) 
         & -11.991985 \pm 0.000089 & -9.398659 \pm 0.000084 \\
      13 & (-1,-10,-100,-10,-1) 
         & -2.914 \pm 0.022 & -0.300 \pm 0.010 \\
      14 & (-1,-100,-10,-100,-1) 
         & -3.237 \pm 0.011 & -0.6648 \pm 0.0028  \\
      15 & (-1,-1,-100,-1,-1) 
         & -12.686 \pm 0.014 & -10.108 \pm 0.010 \\
      16 & (-1,-100,-1,-100,-1) 
         & -14.7067 \pm 0.0077 & -12.1136 \pm 0.0071  \\
      17 & (-1,-100,-100,-1,-1) 
         & -182.32 \pm 0.11 & -179.722 \pm 0.039  \\
      18 & (-1,-100,-10,-100,-10) 
         & -6.3102 \pm 0.0062 & -3.7281 \pm 0.0013  \\
      19 & \left(-1,-\frac{1}{4},-\frac{1}{9},-\frac{1}{16},-\frac{1}{25}\right) 
         & -19.0031 \pm 0.0077 & -16.4136 \pm 0.0021  \\
      20 & \left(-1,-\frac{1}{9},-\frac{1}{4},-\frac{1}{25},-\frac{1}{16}\right) 
         & -15.1839 \pm 0.0046 & -12.5995 \pm 0.0016 \\
      21 & \left(-1,-1,-\frac{1}{4},-1,-1\right) 
         & -9.7628 \pm 0.0028 & -7.17588 \pm 0.00079  \\
      22 & \left(-1,-\frac{1}{4},-\frac{1}{4},-1,-1\right) 
         & -9.5072 \pm 0.0036 & -6.9186 \pm 0.0014  \\
      23 & \left(-1,-\frac{1}{4},-1,-\frac{1}{4},-1\right) 
         & -12.6308 \pm 0.0031 & -10.04241 \pm 0.00083  \\
      24 & \left(-1,-\frac{1}{4},-\frac{1}{9},-\frac{1}{4},-1\right) 
         & -11.0200 \pm 0.0056 & -8.4281 \pm 0.0030 \\
      25 & \left(-1,-\frac{1}{9},-\frac{1}{4},-\frac{1}{9},-1\right) 
         & -19.1966 \pm 0.0070 & -16.6095 \pm 0.0043 \\
     \hline
    \end{array}\)
    \caption{\itshape  $\cO(\epsilon)$  five-point remainders for amplitudes {\rm (}$\mathcal{E}_5^{(2)}${\rm )}  and 
    Wilson loops   {\rm (}$\mathcal{E}_{5,\mathrm{WL}}^{(2)}${\rm )}.}
    \label{tbl:remainders}
  \end{table}

  \begin{table}
  \scriptsize
    \noindent\center\(
    \begin{array}{|c|c|c|c|}
      \hline
      \# & (s_{12},s_{23},s_{34},s_{45},s_{51}) 
         & \mathcal{E}_{5}^{(2)}-\mathcal{E}_{5,\mathrm{WL}}^{(2)}
         & |\mathcal{E}_{5}^{(2)}-\mathcal{E}_{5,\mathrm{WL}}^{(2)} + \frac{5}{2} \zeta_5| / \sigma\\
      \hline
      1 & (-1,-1,-1,-1,-1) 
        & -2.592645 \pm 0.000048 & 6.8 \\
      2 & (-1,-1,-2,-1,-1) 
        & -2.5894 \pm 0.0025 &  1.2 \\
      3 & (-1,-2,-2,-1,-1) 
        & -2.5932 \pm 0.0027 & 0.32 \\
      4 & (-1,-2,-3,-4,-5) 
        & -2.592697 \pm 0.000036 & 10  \\
      5 & (-1,-1,-3,-1,-1) 
        & -2.5933 \pm 0.0028 & 0.35 \\
      6 & (-1,-2,-1,-2,-1) 
        & -2.592697 \pm 0.000025 & 15 \\
      7 & (-1,-3,-3,-1,-1) 
        & -2.5893 \pm 0.0036 & 0.82 \\
      8 & (-1,-2,-3,-2,-1) 
        & -2.5856 \pm 0.0030 & 2.2 \\
      9 & (-1,-3,-2,-5,-4) 
        & -2.5897 \pm 0.0032 & 0.82 \\
      10 & (-1,-3,-1,-3,-1) 
         & -2.5851 \pm 0.0026 & 2.8 \\
      11 & (-1,-4,-8,-16,-32) 
         & -2.5920 \pm 0.0093 & 0.034 \\
      12 & (-1,-8,-4,-32,-16) 
         & -2.59333 \pm 0.00012 & 8.3 \\
      13 & (-1,-10,-100,-10,-1) 
         & -2.614 \pm 0.024 & 0.89 \\
      14 & (-1,-100,-10,-100,-1) 
         & -2.572 \pm 0.011 & 1.9 \\
      15 & (-1,-1,-100,-1,-1) 
         & -2.578 \pm 0.017 & 0.80 \\
      16 & (-1,-100,-1,-100,-1) 
         & -2.593 \pm 0.010 & 0.071 \\
      17 & (-1,-100,-100,-1,-1) 
         & -2.60 \pm 0.11 & 0.039 \\
      18 & (-1,-100,-10,-100,-10) 
         & -2.5820 \pm 0.0063 & 1.6 \\
      19 & \left(-1,-\frac{1}{4},-\frac{1}{9},-\frac{1}{16},-\frac{1}{25}\right) 
         & -2.5894 \pm 0.0080 & 0.36 \\
      20 & \left(-1,-\frac{1}{9},-\frac{1}{4},-\frac{1}{25},-\frac{1}{16}\right) 
         & -2.5844 \pm 0.0049 &  1.6 \\
      21 & \left(-1,-1,-\frac{1}{4},-1,-1\right) 
         & -2.5869 \pm 0.0029 &  1.9 \\
      22 & \left(-1,-\frac{1}{4},-\frac{1}{4},-1,-1\right) 
         & -2.5886 \pm 0.0038 &  0.96 \\
      23 & \left(-1,-\frac{1}{4},-1,-\frac{1}{4},-1\right) 
         & -2.5884 \pm 0.0032 &  1.2 \\
      24 & \left(-1,-\frac{1}{4},-\frac{1}{9},-\frac{1}{4},-1\right) 
         & -2.5919 \pm 0.0064 &  0.064\\
      25 & \left(-1,-\frac{1}{9},-\frac{1}{4},-\frac{1}{9},-1\right) 
         & -2.5870 \pm 0.0082 & 0.65 \\
     \hline
    \end{array}
    \)
    \caption{\itshape Difference of the five-point amplitude and Wilson loop two-loop remainder functions 
             at $\cO(\epsilon)$, and its distance from
             $-\frac{5}{2}\zeta_5\sim -2.592319$ in units of 
             $\sigma$,
the standard deviation reported by the CUBA numerical integration package
\cite{Hahn:2004fe}.}
    \label{tbl:differenceOfRemainders}
  \end{table}

\begin{figure}[h!]
    \center
    \includegraphics[width=0.9 \linewidth]{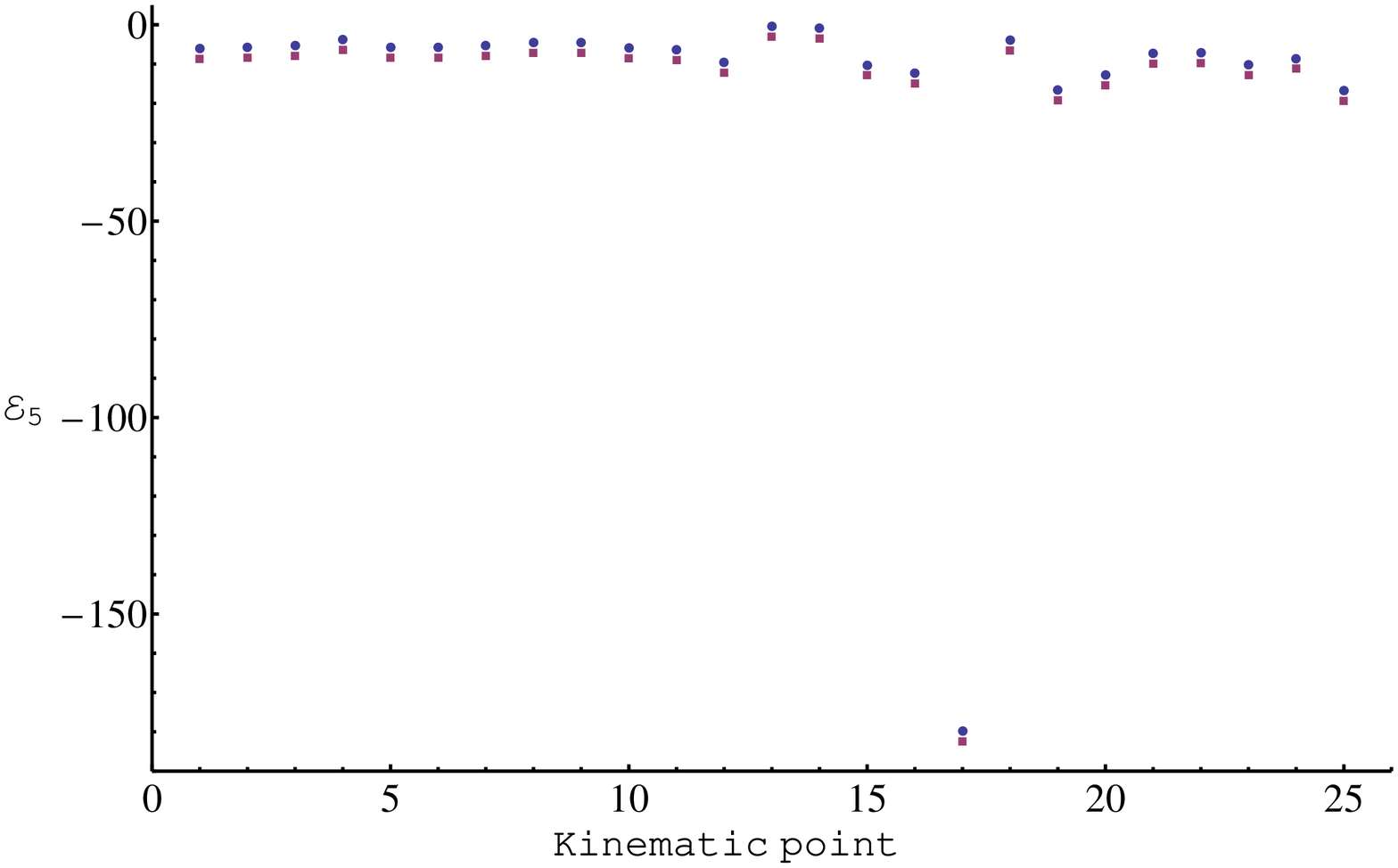}
    \caption{\it Remainder functions at $\cO(\epsilon)$ for the amplitude (circle)
             and the Wilson loop (square).}
    \label{fig:remainders1}
  \end{figure}

\begin{figure}[h!]
    \center
    \includegraphics[width=0.9 \linewidth]{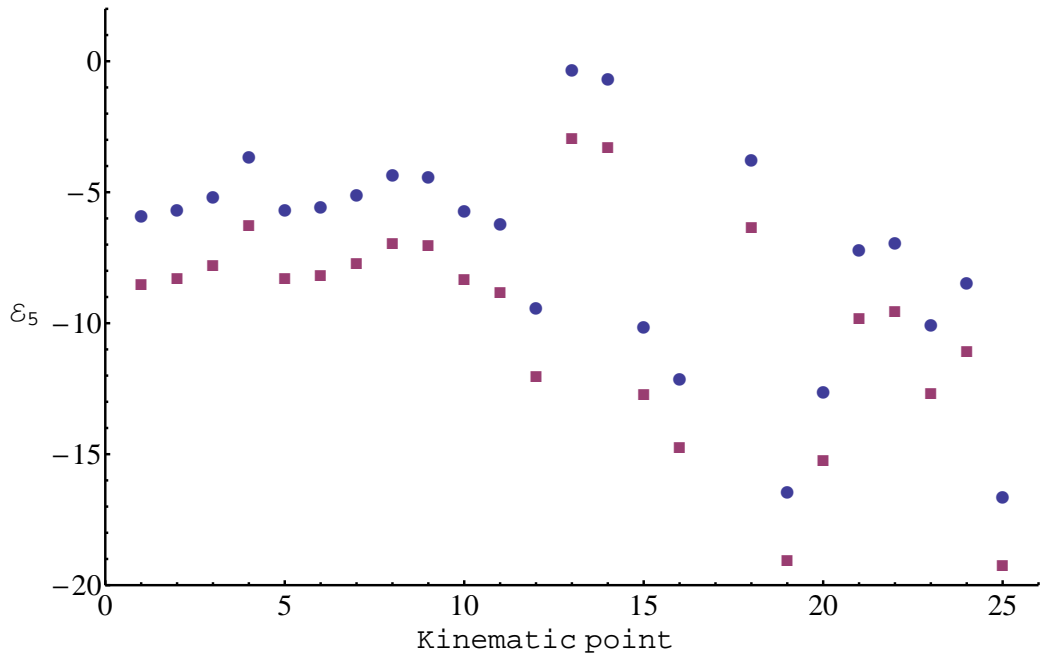}
    \caption{\itshape Remainder functions at $\cO(\epsilon)$ for the amplitude (circle)
             and the Wilson loop (square). In this Figure we have eliminated data point 17 and zoomed in on the others.}
    \label{fig:remainders2}
  \end{figure}

\begin{figure}[h!]
    \center
    \includegraphics[width=0.80 \linewidth]{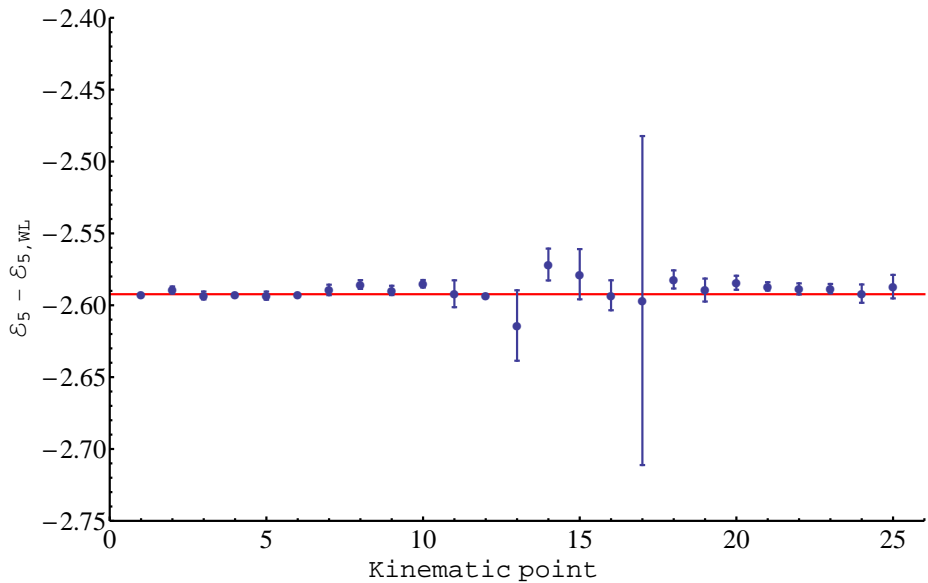}
    \caption{\itshape Difference of the remainder functions 
$\mathcal{E}_{5}^{(2)}-\mathcal{E}_{5,\mathrm{WL}}^{(2)}$.}
    \label{fig:differenceOfRemainders}
  \end{figure}

 We have numerically evaluated both the five-point two-loop amplitude and Wilson loop
  up to $\cO(\epsilon)$   at $25$ Euclidean kinematic points, i.e.~points in the subspace of the kinematic
  invariants with all $s_{ij}<0$. 
  The choice of these points and the values
  of the remainder functions $\mathcal{E}_5^{(2)}$, $\mathcal{E}_{5,\mathrm{WL}}^{(2)}$
  at $\cO(\epsilon)$ together with the errors
reported by the CUBA numerical integration library 
\cite{Hahn:2004fe}
  appear in Table
  \ref{tbl:remainders}, while in Figures \ref{fig:remainders1} and
  \ref{fig:remainders2} we plot both remainders for all kinematic points.
  We have calculated the difference between the amplitude and Wilson loop  remainders,  see
  Table \ref{tbl:differenceOfRemainders} and Figure \ref{fig:differenceOfRemainders}. 
  Remarkably, this difference also appears to be constant (within our numerical precision) as in the four-point case, 
  and hence  we conjecture that
  \begin{equation}
    \mathcal{E}_5^{(2)} = \mathcal{E}_{5,\mathrm{WL}}^{(2)} - \frac{5}{2} \zeta_5
    \ .
  \end{equation}
  It is also intriguing that the constant difference is fit very well by a simple
rational multiple of $\zeta_5$, rather than a linear combination
of $\zeta_5$ and $\zeta_2 \zeta_3$ as would have been allowed
more generally by transcendentality.

 In the last column of Table 
 \ref{tbl:differenceOfRemainders} we give the distance of our results from 
 this conjecture in units of their standard deviation.

  For kinematic points $1$, $4$ and $6$  we have evaluated the remainder functions 
  with
  even higher precision, and  found agreement with the conjecture to 
   $4$   digits. 
   A remark is in order here. 
   By increasing the precision, the mean value of the difference of remainders approaches the conjectured
  value, but we notice that in units of $\sigma$ it drifts away from it, 
   hinting at a potential underestimate of the errors. 
  To test our error estimates we used the remainder functions at $\mathcal{O}( \eps^0 )$, that are known to vanish.
Our analysis confirmed that,  as we increase the desired precision, 
the actual precision of the mean value does increase,
  but on the other hand reported errors  tend to become increasingly
underestimated.

\section*{Acknowledgements}

We would like to thank  M.~Czakon  and V.~V.~Khoze for many helpful discussions. 
MS is grateful to F.~Cachazo for a comment which inspired his interest
in this problem, and to L.~ Dixon and A.~Volovich for  
useful correspondence and discussions.
We would like to thank the QMUL High Throughput Computing Facility, and especially Alex Martin and Christopher Walker, for providing us with the necessary computer power for the high precision numerical evaluations. 
We would also like to thank  Terry Arter  and Alex Owen for technical assistance. 
This work was supported by the STFC under the
Queen Mary Rolling Grant ST/G000565/1 and the IPPP Grant ST/G000905/1, 
by the US Department of Energy under contract DE-FG02-91ER40688, and by the 
US National Science Foundation under grant PHY-0638520. 
GT is supported by an EPSRC Advanced Research Fellowship EP/C544242/1.


\startappendix

\section{Details of the Two-Loop Four-Point Wilson Loop to All Orders in $\epsilon$}
\label{4ptallorders}
In this appendix, we present the results for the separate classes of Wilson loop diagrams contributing to a four-point loop. In all cases (with the exception of the ``hard" diagram) our results are valid to all orders in the dimensional regularization parameter $\eps$.  These expressions are given in terms of hypergeometric functions. We
also expand these up to $\cO(\eps)$ with the help of the mathematica
packages {\tt HPL} and {\tt HypExp}~\cite{HPL,HypExp}.

\subsection{Two-loop Cusp Diagrams}

\begin{figure}[h]
\begin{center}
\scalebox{0.60}{
\fcolorbox{white}{white}{
  \begin{picture}(614,154) (40,-135)
    \SetWidth{0.5}
    \SetColor{Black}
    \ArrowLine(40.01,16.37)(107.31,-135.5)
    \ArrowLine(286.46,19.1)(353.76,-132.77)
    \ArrowLine(491.99,17.28)(559.29,-134.59)
    \ArrowLine(206.44,-50.93)(40.92,16.37)
    \ArrowLine(450.16,-53.66)(286.46,19.1)
    \ArrowLine(654.77,-55.47)(492.9,17.28)
    \Gluon(347.39,-119.13)(350.12,-55.47){6.82}{3.43}
    \Gluon(99.13,-117.31)(101.85,-8.18){6.82}{7.15}
    \Gluon(67.3,-44.56)(186.43,-42.74){6.82}{7.93}
    \Gluon(350.12,-55.47)(411.96,-35.47){6.82}{3.43}
    \Gluon(306.47,-25.46)(350.12,-55.47){6.82}{3.43}
    \Gluon(600.21,-72.75)(631.13,-44.56){6.82}{2.57}
    \SetColor{Blue}
    \Vertex(591.12,-82.76){13.85}
    \SetColor{Black}
    \Gluon(551.1,-114.59)(582.93,-92.76){6.82}{2.57}
  \end{picture}
}
}
\end{center}

\caption{{\it The two-loop cusp corrections. The second diagram
appears with its mirror image where two of the gluon legs of the three-point vertex are attached to the other edge; these two diagrams are equal. The blue bubble in the third diagram represents the gluon self-energy correction calculated in dimensional reduction.  }}
\label{fig:cusps}
\end{figure}
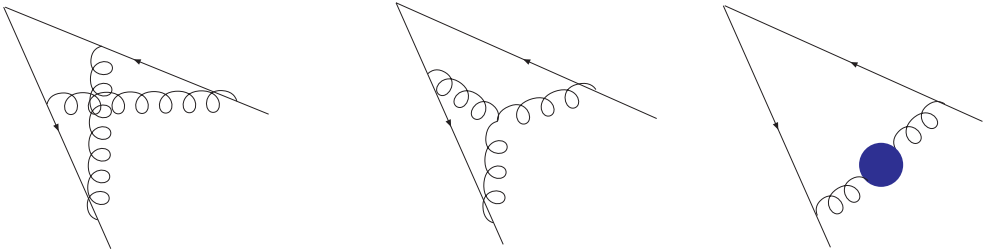

The total contributions of all diagrams that cross a single cusp (the
crossed diagram across a cusp, the self-energy diagram across the cusp
and the vertex across the cusp, as depicted in Figure \ref{fig:cusps}) is easily seen to be%
\footnote{In this and the  following formulae, a  factor of $\cC$ is suppressed in each diagram, where $\cC$ is defined in \eqref{multby}.  } 
\beq
\label{cusps}
( -{s })^{ -2 \eps}\
{1\over  16 \eps^4} \left[ {\Gamma (1 + 2 \eps) \Gamma ( 1 - \eps ) \over
\Gamma ( 1 + \eps )} - 1 \right]
\ . 
\eeq
Adding the contributions for the four cusps, we obtain
\begin{align}
  {\cal W}_{\rm cusp}&\ = \
\Big[ (-s)^{-2\eps}+
  (-t)^{-2\eps}\Big]
  {1\over 8
    \eps^4}\left[{\Gamma(1+2\eps)\Gamma(1-\eps)\over
      \Gamma(1+\eps)}\, -\, 1\right]\\
  &\ =\Big[(-s )^{-2\eps}+(-t )^{-2\eps}\Big]
  \,\nonumber\\
  &\times    \Big[{1\over
    \eps^2}{\pi^2\over 24}\, -\, {1\over
    \eps}{\zeta_3\over 4}\, +\, {\pi^4\over 80}-  \frac{\eps}{12}
  \left(\pi ^2 \zeta_3+9 \zeta_5\right)
+\cO(\eps^2)\Big]\label{wcusp}
    \ .
\end{align}

\subsection{The Curtain Diagram}

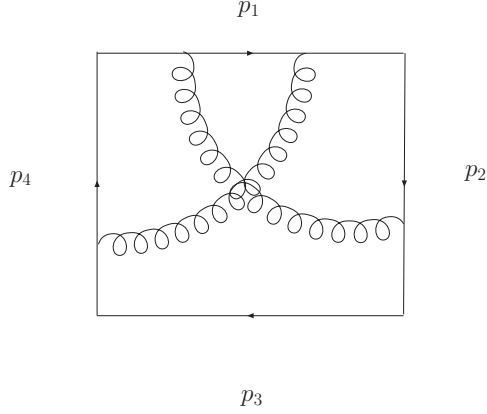
\begin{figure}[t]
\begin{center}
\scalebox{0.55}{
\fcolorbox{white}{white}{
 \begin{picture}(343,284) (166,-48)
   \SetWidth{0.5}
   \SetColor{Black}
   \ArrowLine(225,194)(435,194)
   \ArrowLine(435,14)(225,14)
   \Text(479,107)[lb]{\Large{\Black{$p_2$}}}
   \Text(325,-48)[lb]{\Large{\Black{$p_3$}}}
   \Text(166,103)[lb]{\Large{\Black{$p_4$}}}
   \ArrowLine(436,194)(435,15)
   \ArrowLine(225,14)(225,195)
   \Text(323,220)[lb]{\Large{\Black{$p_1$}}}
   \GlueArc(399.56,187.27)(115.82,176.17,287.82){-7.5}{14.69}
   \GlueArc(234.54,196.2)(133.48,-93.67,-0.95){-7.5}{14}
  \end{picture}
}
}
\end{center}  \caption{\it One of the four curtain diagrams. The remaining three are obtained by cyclic permutations of the momenta.}
\label{curtaindiagram}
\end{figure}

The contribution of all four  curtain diagrams is
\begin{align}
  {\cal W}_{\rm curtain}&=(st)^{-\eps}\left( - {1\over 2
    \eps^4}\right) 
    \left[1-{\Gamma(1-\eps)^2\over
      \Gamma(1-2\eps)}\right]\\ \nonumber 
  &=\Big[(-s)^{-2\eps}+(-t)^{-2\eps}\Big]\
\bigg[   - {1\over
    \eps^2}{\pi^2\over 24}\, -\, {1\over
    \eps}{\zeta_3\over 2}\, -\, {\pi^4\over 160}\, +\, {\pi^2\over 48}\log^2
  x\\
& -\eps \Big(-\frac{1}{4} \zeta_3 \log ^2 x +\frac{3 
  }{2} \zeta_5 -\frac{\pi^2}{12} \,  \zeta_3\Big)+
\cO(\eps^2)\bigg]\nonumber
\ . 
\end{align}

\subsection{The Factorised Cross Diagram}

The factorised cross diagram is given by the product of two finite
one-loop Wilson loop diagrams, expressed each by~(\ref{4pnt1loop})
\beqa
\nonumber 
&&\hspace{-1cm} {1\over 2 \eps^2} \left( 
{st\over u}\right)^{-\eps} 
\Big[x^{-\eps }{}_2 {\rm F}_1( \eps, \eps ,1+\eps,-1/x)+x^{\eps} {}_2
{\rm F}_1(\eps, \eps,1+\eps,-x) 
  - 2\pi \eps \cot (\eps \pi)\Big]
\ . \label{fcross}
\eeqa
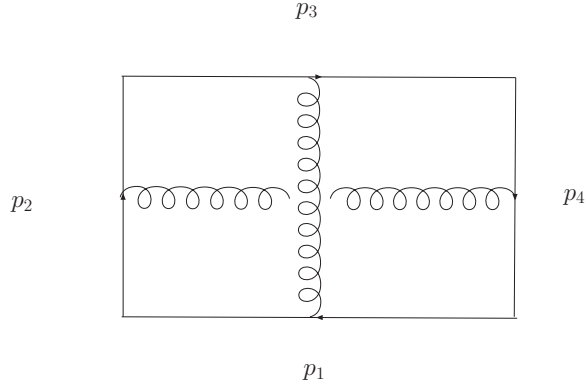
\begin{figure}[t]
\begin{center}
\scalebox{0.55}{
\fcolorbox{white}{white}{
 \begin{picture}(410,265) (150,-108)
   \SetWidth{0.5}
   \SetColor{Black}
   \Gluon(353,100)(354,-64){7.5}{10.29}
   \Gluon(224,18)(340,17){7.5}{6.86}
   \Gluon(368,17)(495,17){7.5}{7.64}
   \ArrowLine(496,-65)(225,-64)
   \ArrowLine(225,101)(495,100)
   \ArrowLine(495,100)(494,-64)
   \ArrowLine(226,-64)(226,101)
   \Text(346,141)[lb]{\Large{\Black{$p_3$}}}
   \Text(351,-108)[lb]{\Large{\Black{$p_1$}}}
   \Text(530,13)[lb]{\Large{\Black{$p_4$}}}
   \Text(150,8)[lb]{\Large{\Black{$p_2$}}}
 \end{picture}
}
}
\end{center}  \caption{\it One of the four factorised cross diagram.}
\label{factcrossdiagram}
\end{figure}
The result for the factorised cross is therefore
\beqa
\nonumber
 && \hspace{-1.2cm}
- {1\over 8\eps^4}\left( st\over u\right)^{-2\eps}  \bigg[\, _ 2 {\rm F}_1
         (\eps , \eps ;
         1 + \eps ; -x) x^{\eps } + \, _ 2 {\rm F}_1
       \left (\eps
           , \eps ;
        1 + \eps ; -\frac {1} {x} \right) x^{-\eps 
        }
        -2\pi
      \eps \cot(\eps \pi) \bigg]^2 
      \nonumber \\
&=&\Big[(-s)^{-2\eps}+(-t)^{-2\eps}\Big]
\times\Big(g_0 + g_{-1} \eps+ \cO(\eps^2)\Big)
\ , 
\eeqa
with
\beqa
g_0&=&-\frac{1}{64} \left(\log ^2 x+\pi ^2\right)^2
\ , 
\\ 
g_{-1}&=&
\frac{1}{192} \Big(\log ^2 x+\pi ^2\Big) \Big[\log ^3 x \, -\, 6
   \log (x+1) \log ^2 x -12 \text{Li}_2(-x) \log x
   \ , 
   \nonumber \\
&+& 3 \pi ^2
   \log  x -6 \pi ^2 \log (x+1)+12 \, \text{Li}_3(-x)-12 \zeta_3
   \Big]
   \ . 
\label{gm1}
\eeqa

\subsection{The Y  Diagram }

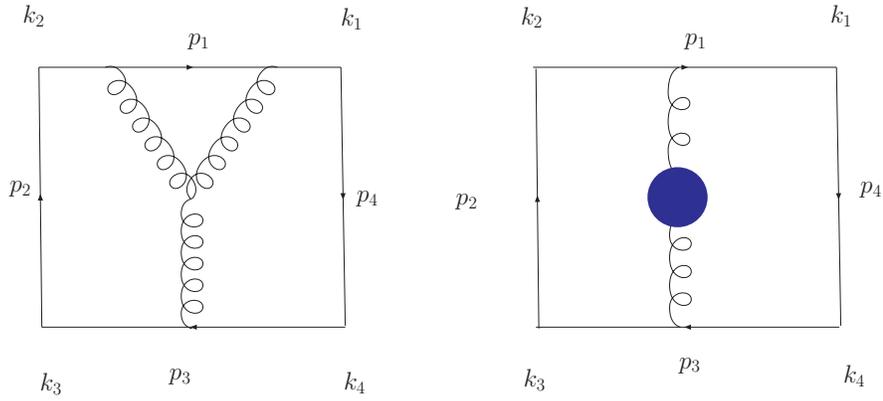
\begin{figure}[t]
\begin{center}
\scalebox{0.55}{
\fcolorbox{white}{white}{
 \begin{picture}(614,269) (40,-44)
   \SetWidth{0.5}
   \SetColor{Black}
   \Gluon(105.92,180.17)(165.32,90.09){7.42}{6.06}
   \Gluon(164.33,90.09)(223.73,180.17){7.42}{6.06}
   \Gluon(165.32,0.99)(165.32,90.09){7.42}{5}
   \Text(163.34,193.04)[lb]{\Large{\Black{$p_1$}}}
   \ArrowLine(270.26,1.98)(62.37,1.98)
   \ArrowLine(60.39,180.17)(267.29,180.17)
   \ArrowLine(398.95,180.17)(606.84,180.17)
   \ArrowLine(608.82,1.98)(400.93,1.98)
   \Gluon(499.93,1.98)(499.93,78.21){7.42}{3.49}
   \Gluon(498.94,105.92)(498.94,180.17){7.42}{2.57}
   \SetColor{Blue}
   \Vertex(497.95,91.08){20.6}
   \Text(504.88,193.04)[lb]{\Large{\Black{$p_1$}}}
   \Text(150.47,-37.62)[lb]{\Large{\Black{$p_3$}}}
   \Text(40.59,92.07)[lb]{\Large{\Black{$p_2$}}}
   \Text(500.92,-29.7)[lb]{\Large{\Black{$p_3$}}}
   \Text(625.65,91.08)[lb]{\Large{\Black{$p_4$}}}
   \SetColor{Black}
   \ArrowLine(402.91,0.99)(400.93,179.18)
   \ArrowLine(62.37,1.98)(60.39,180.17)
   \ArrowLine(267.29,180.17)(270.26,2.97)
   \Text(279.17,85.14)[lb]{\Large{\Black{$p_4$}}}
   \Text(347.47,83.16)[lb]{\Large{\Black{$p_2$}}}
   \Text(61.38,-44.55)[lb]{\Large{\Black{$k_3$}}}
   \Text(270.26,-43.56)[lb]{\Large{\Black{$k_4$}}}
   \Text(49.5,208.88)[lb]{\Large{\Black{$k_2$}}}
   \Text(268.28,206.9)[lb]{\Large{\Black{$k_1$}}}
   \Text(392.02,207.89)[lb]{\Large{\Black{$k_2$}}}
   \Text(394,-41.58)[lb]{\Large{\Black{$k_3$}}}
   \Text(613.77,-38.61)[lb]{\Large{\Black{$k_4$}}}
   \Text(604.86,208.88)[lb]{\Large{\Black{$k_1$}}}
   \ArrowLine(606.84,180.17)(609.81,2.97)
 \end{picture}
}
}
\end{center}  \caption{\it The Y diagram together with the self-energy diagram. The sum of these two topologies gives a maximally transcendental contribution.}
\label{Ydiagram}
\end{figure}

The diagrams in Figure \ref{Ydiagram}  correspond to the following contribution to the 
two-loop Wilson loop:
\begin{align}
\label{vertex}
{\cal W}_{\rm Y} &\ = \ - {1\over
   16\eps}\, {\Gamma(1+2\eps)\over
   \Gamma(1+\eps)^2}\, \, u\,  
   \\
 &\times \int_0^1\!d\tau_1 \int_0^1\!d\tau_2\Big[
2I(z_1(\tau_1),z_2(\tau_2),z_2(\tau_2)) -
I(z_1(\tau_1),z_2(\tau_2),k_1 )- I(z_1(\tau_1),z_2(\tau_2),k_2)\Big]
\ , 
\nonumber 
\end{align}
where
$z_1(\tau_1)=k_3-p_3 \tau_1,\ z_2(\tau_2)=k_1 -p_1 \tau_2$,
and 
\beq
    I(z_1,z_2,z_3)\ = \ \int_0^1\!d\sigma(\sigma(1-\sigma))^{\eps}\Big[-(-z_1+\sigma z_2+(1-\sigma)z_3)^2+i0\Big]^{-1-2\eps}
    \ , 
\eeq
where  $z_2, z_3$ must
be lightlike. The evaluation of \eqref{vertex} gives
\beqa
{\cal W}_{\rm Y} & = &  
\left( s t\over u\right)^{-2\eps} {1\over 64\eps^4}\, 
\Big[-2(x+1)^{-2 \eps }  \frac{\Gamma (1+2 \eps ) \Gamma (-\eps +1) }{\Gamma (1+\eps )}
\nonumber \\ 
&+& 4x^{\eps }\frac{ \Gamma (-\eps +1)^2}{\Gamma (-2 \eps +1)} \, _2{\rm F}_1(\eps ,1+2
   \eps ;1+\eps ;-x) 
   \nonumber  
+x^{2 \eps }\, _2{\rm F}_1\left(2 \eps ,2 \eps ;1+2 \eps ;-x\right)
 \nonumber  \\
&-&4 \pi
    \eps  \cot (2 \pi  \eps )+\Gamma (1+2 \eps ) 
    \Gamma (-2   \eps +1)\ +\  x\leftrightarrow {1\over x} 
 \Big] 
 \nonumber
\eeqa
We multiply by four to obtain the contribution of all such diagrams. Then
the expansion of this contribution in $\eps$ begins at $\cO(\eps^{-1})$, 
\begin{align}
&\Big[(-s)^{-2\eps}+(-t)^{-2\eps}\Big]
\times\Big[{c_1\over \eps} +c_0 +c_{-1} \eps+ \cO(\eps^2)\Big]
\ , 
\end{align}
where
\begin{align}
  c_1\ =\ & - 
  \frac{1}{48} \Big[\log ^3 x -6 \log (x+1) \log ^2 x -12
    \text{Li}_2(-x) \log  x +3 \pi ^2 \log  x )
    \nonumber\\ 
    & -6 \pi ^2 \log (x+1
+12 \text{Li}_3(-x)-12 \zeta_3\Big] 
\, , 
\label{c1}\\[15pt]
c_0\ =\ &\frac{1}{960} \Big[5 \log ^4 x -40 \log (x+1) \log ^3 x +120
\log ^2(x+1) \log ^2 x +10 \pi ^2 \log ^2 x 
\nonumber\\
&-120 \pi ^2 \log (x+1) \log x +480 \log (x+1) \text{Li}_2(-x) \log  x -240\,  \text{Li}_3(-x) \log  x 
    \nonumber\\
   & +480 S_{1,2}(-x) \log  x 
  -240 \zeta_3 \log x+120 \pi ^2 \log ^2(x+1)-480\,  \log (x+1) \text{Li}_3(-x)
   \nonumber\\
&   +480
   \text{Li}_4(-x)
 -480 S_{2,2}(-x)+480 \log (x+1) \zeta_3 +\pi ^4
\Big]
\, , 
\end{align}
\begin{align}
c_{-1}=&\frac{1}{240} \Big[-2 \log^5 x +10 \log (x+1) \log^4 x +10
\log ^2(x+1) \log^3 x +20 \text{Li}_2(-x) \log^3 x \nonumber\\
& -5 \pi ^2
   \log^3 x -20 \log^3(x+1) \log^2 x -30 \zeta_3 \log^2 x +30 \pi^2 \log^2(x+1) \log x\nonumber\\
&-120 \log^2(x+1)
   \text{Li}_2(-x) \log  x +120 \log (x+1) \text{Li}_3(-x) \log  x -120 \text{Li}_4(-x) \log  x \nonumber\\
&+120 S_{2,2}(-x) \log (x)-240
   \log (x+1) S_{1,2}(-x) \log  x -240 S_{1,3}(-x) \log  x \nonumber\\
&+120 \log (x+1) \zeta_3 \log  x +4 \pi ^4 \log  x -20 \pi ^2
   \log^3(x+1)\nonumber\\
&-8 \pi^4 \log (x+1)+120 \log^2(x+1) \text{Li}_3(-x)\nonumber\\
&-40 \text{Li}_2(-x) \text{Li}_3(-x)-240 \log (x+1)
   \text{Li}_4(-x)+240 \text{Li}_5(-x)+240 \log (x+1) S_{2,2}(-x)\nonumber\\
&+40 H_{2,3}(-x)+120 H_{3,2}(-x)+40 \text{Li}_2(-x)
   S_{1,2}(-x)-40 H_{2,1,2}(-x)\nonumber\\
&-120 H_{2,2,1}(-x)-240 \zeta_5-120
   \log^2(x+1) \zeta_3-30 \pi^2 \zeta_3\Big]\label{cm1}
   \ . 
\end{align}

\subsection{The Half-Curtain Diagram}

We now consider the ``half-curtain" diagram, 
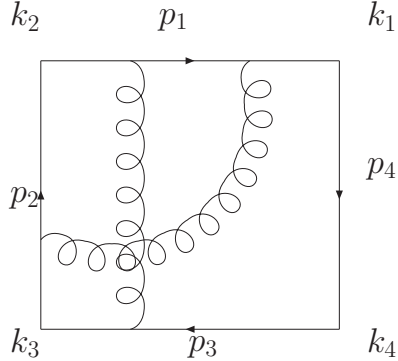
\begin{figure}[h]
\begin{center}
\scalebox{.75}{
\fcolorbox{white}{white}{
  \begin{picture}(210,181) (120,-239)
    \SetWidth{0.5}
    \SetColor{Black}
    \ArrowLine(135,-224)(135,-89)
    \ArrowLine(135,-89)(285,-89)
    \ArrowLine(285,-224)(135,-224)
    \ArrowLine(285,-89)(285,-224)
    \Text(120,-164)[lb]{\Large{\Black{$p_2$}}}
    \Text(195,-74)[lb]{\Large{\Black{$p_1$}}}
    \Text(300,-74)[lb]{\Large{\Black{$k_1$}}}
    \Text(300,-149)[lb]{\Large{\Black{$p_4$}}}
    \Text(300,-239)[lb]{\Large{\Black{$k_4$}}}
    \Text(210,-239)[lb]{\Large{\Black{$p_3$}}}
    \Text(120,-239)[lb]{\Large{\Black{$k_3$}}}
    \Text(120,-74)[lb]{\Large{\Black{$k_2$}}}
    \Gluon(180,-89)(180,-224){7}{7.71}
    \GlueArc(170.43,-114.09)(73.95,-118.63,19.83){-7}{11.14}
  \end{picture}
}
}
\end{center}\caption{\it Diagram of the half-curtain topology.}
\end{figure}
whose contribution to the Wilson loop is
\begin{align}
{\cal  W}_{\rm hc}(x)=& -{1\over 8}\int_0^1\!\!d\sigma \, d\rho \, d\tau_1\, 
\int_{\tau_1}^1d\tau_2 {s\over
    (-s\sigma\tau_2)^{1+\eps}} {u\over (-s \tau_1 -t \rho -u \rho
    \tau_1)^{1+\eps}}
    \nonumber \\
=&\, \, {1\over 8}\left(st\over u\right)^{-2\eps}\!\!
(1+x)^{-\eps}
\int_{0}^{1}\!d\sigma
\int_1^{-1/x}\!da\int_1^{-x}\!db\int_{1-b\over 1+x}^1\!d \tau_2 {1\over
  (\sigma \tau_2)^{1+\eps}}{1\over (1-a b)^{1+\eps}}
  \, , 
\end{align}
where we have changed variables in the second line to 
$a=1+u\rho/  s$, and
  $ b=1+u\tau_1/  t$. 
  
The evaluation of this diagram  and of that with $s\leftrightarrow t$ leads to 
\begin{align}
\label{hcu}
&{\cal W}_{\rm hc}(x)+{\cal W}_{\rm hc}(1/x) =   {1\over 8\eps^4}\left(st\over u\right)^{-2\eps}\
\nonumber\\
& \times \Big[ 2\pi\eps\cot(2\pi\eps)-{1\over
  2}x^{-2\eps}\,{}_2{\rm F}_1(2\eps,2\eps;1+2\eps;-1/x)
  \nonumber \\
  &
  -{1\over 2}x^{2\eps}\,{}_2{\rm F}_1(2\eps,2\eps;1+2\eps;-x)\nonumber\\
& +\Big[(1+x)^{-\eps}+(1+1/x)^{-\eps}\Big]\Big(x^{-\eps}\,{}_2{\rm F}_1(\eps,\eps;1+\eps;-1/x)
\nonumber \\ 
&  +x^{\eps}\,{}_2{\rm F}_1(\eps,\eps;1+\eps;-x)
-2\pi\eps\cot(\pi\eps)\Big)\Big]
\ . 
\end{align}
This can be expanded in $\eps$ using
\beqa
  {}_2{\rm F}_1(\eps,\eps;1+\eps;x)&\!\!=\!\!&1+\eps^2{\rm Li}_2(x)-\eps^3\big[ {\rm Li}_3(x)-S_{12}(x)\big]
  \nonumber \\ 
  &\!\!+\!\!&
  \eps^4\big[{\rm Li}_4(x)-S_{22}(x)+S_{13}(x)\big]+\cO(\eps^6)\ ,
\eeqa
The contribution of all diagrams of the half-curtain type  
is obtained by multiplying \eqref{hcu} by a factor of four. 
One obtains thus
\begin{align}
&\Big[(-s)^{-2\eps}+(-t)^{-2\eps}\Big]
\times\ \Big[ {d_1\over \eps} +d_0 +d_{-1}\eps+\cO(\eps^2)\Big]
\ , 
\end{align}
where
\beqa
d_1 &=&  2c_1\ , \\
d_0 &=&  -3c_0-\frac{1}{64} \left(3 \pi ^2-\log ^2 x \right) \left(\log
  ^2 x +\pi ^2\right)\ , \\
d_{-1} &=& {7\over 2}c_{-1} 
\\
      & -&\frac{1}{96} 
      \Big[-\log ^5(x)+6 \log
  (x+1) \log ^4 x +12 \text{Li}_2(-x) \log ^3(x)-3 \pi ^2 \log ^3 x
  \nonumber\\
  &+&6
  \pi ^2 \log (x+1) \log ^2 x -12 
   \text{Li}_3(-x) \log ^2 x -30 \zeta_3 \log ^2 x -42 \pi ^2 \zeta_3\Big]
   \ , 
   \nonumber
\eeqa
and $c_j$ are the coefficients for the ${\rm Y}$ diagram, given
in \eqref{c1}--\eqref{cm1}.

\subsection{The Cross Diagram}
\begin{figure}[h]
\begin{center}
\scalebox{.55}{
\fcolorbox{white}{white}{
 \begin{picture}(343,284) (166,-48)
   \SetWidth{0.5}
   \SetColor{Black}
   \ArrowLine(225,194)(435,194)
   \ArrowLine(435,14)(225,14)
   \Gluon(270,194)(375,14){7.5}{13.46}
   \Gluon(375,194)(270,14){7.5}{13.46}
   \Text(479,107)[lb]{\Large{\Black{$p_2$}}}
   \Text(325,-48)[lb]{\Large{\Black{$p_3$}}}
   \Text(166,103)[lb]{\Large{\Black{$p_4$}}}
   \ArrowLine(436,194)(435,15)
   \ArrowLine(225,14)(225,195)
   \Text(323,220)[lb]{\Large{\Black{$p_1$}}}
 \end{picture}
}
}
\end{center}\caption{\it One of the cross diagrams. As before, the remaining three can be generated by cyclic permutations of the momentum labels. }
\label{crossdiagram}
\end{figure}
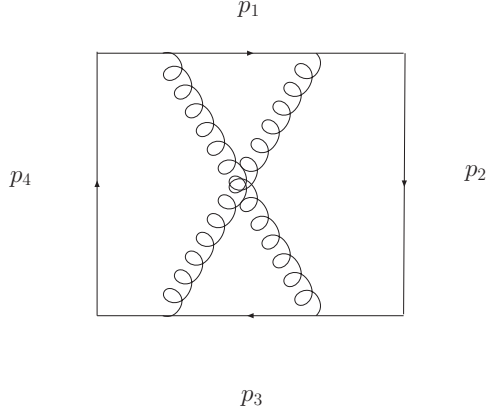

We now consider the cross diagram, whose expression is given by 
\begin{align}
{\cal  W}_{\rm cr}(x)=& -{1\over 8}\int_0^1\!\!d\tau_1\, d\sigma_1\int_0^{\tau_1}\!\!d\tau_2 \int_0^{\sigma_1} \tau_2
 {u\over  (-s \sigma_1 - t \tau_2 -u \sigma_1    \tau_2)^{1+\eps}
   } {u\over  (-s \sigma_2 - t \tau_1 -u \sigma_2    \tau_1)^{1+\eps}}
    \nonumber\\
=&-{1\over 8}\left(st\over u\right)^{-2\eps}
\int_{-1/x}^1\!\!\!\!da_1 
\int_{-x}^{1}\!\!db_1\int_1^{a_1}\!\!da_2\int_{1}^{b_1}\!\!d b_2 {1\over
  (1-a_1b_2)^{1+\eps}}{1\over (1-a_2 b_1)^{1+\eps}}
 \ , 
\end{align}
for which we find 
\begin{align}   {\cal  W}_{\rm cr}(x)
&=-{1\over 8 \eps^4}
   \left(st\over u\right)^{-2\eps}
   \nonumber\\
&
\Big\{
-\frac {1}{4} \, _ 3 {\rm F}_
     2 (2 \eps , 2 \eps ,2 \eps ; 1 +2 \eps ,
      1 + 2 \eps ; -x) x^{2 \eps}
\nonumber \\
&
+ \frac {\Gamma (-\eps + 1)^2 \, _ 3 {\rm F}_2
     (2 \eps , \eps , \eps ; 1 + \eps ,
     1 + \eps ; -x) x^{\eps
         }} { \Gamma (-2 \eps +
      1)} 
      \nonumber \\
 & + {1\over 4}\bigg[\, _ 2 {\rm F}_1
         (\eps , \eps ;
         1 + \eps ; -x) x^{\eps } + \, _ 2 {\rm F}_1
       \Big(\eps
           , \eps ;
        1 + \eps ; -\frac {1} {x} \Big) x^{-\eps } 
        \nonumber \\ 
        &-2\pi
      \eps \cot(\eps \pi) \bigg]^2 
 - \eps^2\pi  \cot (2 \pi  \eps )
   \left(\psi(2 \eps )+\gamma \right)-\pi^2 \eps^2 \cot(\pi \eps)^2\nonumber \\
&  + \Big(x\leftrightarrow {1\over x} \Big)
\Big\}\ . 
\end{align}
Notice the presence of the one-loop finite diagram squared.
Expanding this and multiplying by a factor of two to account for all diagrams  leads to the following result,  
\begin{align}
   {\cal  W}_{\rm cr}(x)&\ =\ \Big[(-s)^{-2\eps}+(-t)^{-2\eps}\Big]
\times\ \Big[f_0 +f_{-1} \eps+ \cO(\eps^2)\Big]
\ , 
\end{align}
where
\begin{align}
  f_0\ =\ &-{1\over 192}(\pi^2 + \log^2 x)^2
  \ , 
  \\[10pt]
f_{-1}\ =  \ &-\frac1{2880} \Big[-3 \log^5 x+30 \log (x+1) \log^4 x+180 \text{Li}_2(-x)   \log^3 x\nonumber\\
& 
-20 \pi ^2 \log^3 x+60 \pi ^2 \log (x+1) \log^2 x-900 \text{Li}_3(-x) \log^2 x
-540 \, \zeta_3 \log^2 x
   \nonumber\\ &
   +180 \pi^2 \text{Li}_2(-x) \log x +2880 \text{Li}_4(-x)
   \log x 
   \nonumber\\
& 
-49 \pi ^4 \log x +30 \pi ^4 \log (x+1)-420 \pi ^2
   \text{Li}_3(-x)-4320 \, \text{Li}_5(-x)\nonumber\\
& 
+4320 \zeta_5-1020 \, \pi^2 \zeta_3\Big]
   \ . 
\end{align}

\subsection{The Hard Diagram}
\label{harddiagram}

The generic $n$-point hard diagram topology is depicted in Figure \ref{generalharddiagram}. In the four-point case, 
one diagram is obtained from  Figure \ref{generalharddiagram} by simply setting  $Q_2 = Q_3 = 0$, and $Q_1 = p_4$. 
There are four such diagrams, obtained by cyclic rearrangements of the momenta. 
We have evaluated the hard diagrams using Mellin-Barnes, arriving at the following result: 
\begin{align}
  W_{\rm hard} \ =  \Big[(-s)^{-2\eps}+(-t)^{-2\eps}\Big]
\times\ \Big[ {h_2\over \eps^2}\, +\, {h_1\over \eps} \, +\, h_0 \, + \, h_{-1}\eps+\cO(\eps^2)\Big]
\ , 
\end{align}
where
\begin{align}
h_2 & ={\pi^2\over 48}
\ , 
\\
h_1&=   - \Big( c_1+{\zeta_3\over 8} \Big) \ , 
\\
\label{h0}
h_0 &\!=\!  -2c_0-\frac{1}{192}  \left(\log^2 x +\pi ^2\right)^2-\frac{1}{48}  \pi^2\left(\log^2 x+\pi ^2\right)+{31\over 1440}\pi^4
\end{align}
\begin{align}
h_{-1}& =  
- \frac{1}{480} \Big[200 H_{2,3} (-x)+
600
 H_{ 3,2}( -x)-200 H_{2,1,2} (-x)-600
  H_{2,2,1} (-x)\nonumber \\ 
  &+ 200 \zeta_3 \text{Li}_2(-x)
  - 200
  \text{Li}_2(-x) \text{Li}_3(-x)
 -200 \text{Li}_2(-x)
  \text{Li}_3(x+1)+1320 \text{Li}_5(-x)
   \nonumber \\
  &  + 20 \text{Li}_2(-x) \log^3(x)+
  60 \text{Li}_3(-x) \log ^2(x)-600 \text{Li}_2(-x) 
  \log^2(x+1) \log x 
  \nonumber \\
 & - 600 \text{Li}_2(x+1) \log ^2(x+1) \log x +100
  \text{Li}_2(-x) \log (-x) \log^2(x+1)
  \nonumber \\
  &+ 600 \text{Li}_3(-x) \log^2(x+1)+20 \pi ^2 \text{Li}_2(-x) \log (x)+600 \text{Li}_3(-x) \log
  (x+1) \log (x)
     \nonumber \\
&  - 600 \text{Li}_4(-x) \log (x)+1200 \text{Li}_4(x+1)
  \log x +200 \text{Li}_2(-x) \text{Li}_2(x+1) \log (x+1)
  \nonumber \\
  &  -  1200
  \text{Li}_4(-x) \log (x+1)+600 S_{2,2}(-x) \log x+1200
  S_{2,2}(-x) \log (x+1)
  \nonumber \\
 & - 240 \zeta_3 \log^2 x -600 \zeta_3 \log^2(x+1)-600 \zeta_3 \log (x+1) \log x -2 \log^5 x 
 \nonumber \\
 &+ 
  5 \log (x+1)
  \log^4 x-400 \log (-x) \log^3(x+1) \log x -100 \pi^2 \log^3(x+1)
  \nonumber \\ 
  &- 40 \pi^2 \log (x+1) \log^2 x +150 \pi^2 \log ^2(x+1)
  \log x +50 \log^2(x+1) \log^3 x
  \nonumber \\
  &- 
  100 \log^3(x+1) \log^2  x
  +7 \pi^4 \log x -35 \pi^4 \log (x+1)-1020 \zeta_5-320 \pi^2
  \zeta_3\Big] \, . 
  \nonumber \\
  \end{align}
  The analytical evaluation of this diagram up to $\cO(\eps^0)$ was obtained in \cite{dhks4}. 
Our evaluation of the  $\cO (\eps^0 )$  terms agrees precisely with that of  \cite{Anastasiou:2009kn} (and with \cite{dhks4} up to a constant term). The evaluation of the $\cO (\eps)$ term  is new and has been performed numerically. 
We have then compared the entire Wilson loop expansion 
to the analytic expression for the amplitude remainder given in 
\eqref{fourpointremainderamplitude}, finding the relation \eqref{diff4ofremainders}.


\section{The One-Loop Wilson Loop Reloaded}

\label{appb}

In this section  we derive a new expression for the all-orders in
$\epsilon$ one-loop finite
Wilson loop diagrams, and hence also a new expression for the all orders in
$\epsilon$ (finite part of
the)  $2me$ box function. We also improve a previous expression for use
in all kinematical regimes.

A general one-loop Wilson loop diagram is given by the following
integral
%
\begin{equation}
  \begin{array}{rcl}
\scalebox{.4}{\fcolorbox{white}{white}{
  \begin{picture}(165,75) (3,-250)
    \SetWidth{0.5}
    \SetColor{Black}
    \Gluon(76,-293)(76,-203){3.5}{11}
    \SetWidth{1.0}
    \DashArrowLine(22,-294)(22,-202){4}
    \DashArrowLine(128,-203)(128,-295){4}
    \SetWidth{1.5}
    \ArrowLine(23,-203)(128,-203)
    \ArrowLine(128,-295)(23,-295)
    \Text(138,-252)[lb]{\Large{\Black{$Q$}}}
    \Text(61,-187)[lb]{\Large{\Black{$p$}}}
    \Text(66,-318)[lb]{\Large{\Black{$q$}}}
    \Text(3,-254)[lb]{\Large{\Black{$P$}}}
  \end{picture}
}
} &:=& {\Gamma ( 1 + \e) e^{  \epsilon \gamma
   }}  \times {\cal F}_\epsilon(s,t,P^2,Q^2)\\[10pt]
&=& {\Gamma ( 1 + \e) e^{  \epsilon \gamma
   }}  \times \int_0^1 d\tau   \int_0^1 d\sigma {u/2 \over \left\{ -
    [P^2+\sigma(s-P^2) +\tau(t-Q^2 +\sigma \tau u)]-i \vare
  \right\}^{1+\epsilon}}\ .
\end{array}
\end{equation}
 Here we have defined $s=(p +P)^2$, $t=(p+Q)^2$ and $u=P^2
+Q^2 -s -t$. 
The relation between this Wilson loop diagram and the corresponding $2me$ box function is~\cite{bht} 
\begin{align}
{\rm finite\  part\  of\  } \left(\scalebox{.2}
{
\fcolorbox{white}{white}{
  \begin{picture}(249,110) (25,-170)
    \SetWidth{1.5}
    \SetColor{Black}
    \Line(95,-123)(200,-123)
    \DashLine(94,-214)(94,-122){0.2}
    \DashLine(200,-123)(200,-215){0.2}
    \Line(200,-215)(95,-215)
    \Text(244,-81)[lb]{\Huge{\Black{$Q$}}}
    \Text(35,-281)[lb]{\Huge{\Black{$P$}}}
    \Text(25,-74)[lb]{\Huge{\Black{$p$}}}
    \Text(243,-280)[lb]{\Huge{\Black{$q$}}}
    \Line(92,-215)(62,-248)
    \Line(93,-215)(79,-257)
    \Line(94,-214)(50,-231)
    \Line(94,-214)(50,-231)
    \Line(246,-105)(202,-122)
    \Line(230,-89)(200,-122)
    \Line(213,-78)(199,-120)
    \Line(202,-215)(230,-246)
    \Line(94,-123)(61,-88)
  \end{picture}
}
}
\right)
 = e^{\epsilon \gamma} \frac{\Gamma(1 + \epsilon) \Gamma^2(1 -
   \epsilon)}{\Gamma(1 - 2 \epsilon)} \times {\cal
   F}_\epsilon(s,t,P^2,Q^2)\ .
\end{align}
Notice that we have included an infinitesimal negative
imaginary part $-i\vare$
in the denominator which dictates the analytic properties of
the integral. This has the opposite sign to the one expected
from a propagator term in a Wilson loop in configuration
space. On the other hand it has the correct sign for the present
application, namely 
for the duality with amplitudes~\cite{Georgiou:2009mp}. One simple way
to deal with this is simply to add an identical positive imaginary
part to all kinematical invariants 
\begin{align}\label{an}
  s\rightarrow s+i \vare, \quad t\rightarrow t+i \vare, \quad P^2\rightarrow P^2+i
  \vare, \quad Q^2\rightarrow Q^2+i \vare\ .
\end{align}
We will assume this in the following.

Changing variables to $\sigma'=\sigma -(P^2-t)/u$ and  $\tau'=\tau
-(P^2-s)/u$ and then dropping the primes, this becomes
\begin{align}
{\cal F}_\epsilon(s,t,P^2,Q^2)=& \int_{t-P^2\over u}^{Q^2-s \over u}d \sigma \int_{s-P^2 \over
    u}^{Q^2-t \over u}d\tau {u/2 \over \left[ - (a^{-1}+\sigma \tau u)\right]^{1+\epsilon}}\\
=&-{1 \over 2\epsilon} \int_{t-P^2\over u}^{Q^2-s \over u}{d \sigma\over \sigma} \left. {1 \over \left[ - (a^{-1}+\sigma \tau u)\right]^{\epsilon}}\right|^{\tau=(Q^2-t)/u}_{\tau=(s-P^2)/u}\ ,
\end{align}
where $a=u/(P^2 Q^2 -s t)$. Note that the analytic continuation of
$a$ implied by~(\ref{an}) is $a\rightarrow a-i\vare$.   Now we split
the integration into two parts, 
\begin{align}
  \int_{t-P^2\over u}^{Q^2-s \over u} =-\int_0^{t-P^2\over u}+\int_{0}^{Q^2-s \over u}
  \ , 
\end{align}
and rescale the integration variable so that it runs between 0 and 1
in each case. It is important to split the integral in this way since
there is a singularity at $\sigma=0$ which one must be very careful
when  integrating over. We obtain in this way 
\begin{align}
{\cal F}_\epsilon(s,t,P^2,Q^2)=&{(-a)^\epsilon \over 2 \epsilon} \int_{0}^{1}{d \sigma\over \sigma} 
\left\{ 
{1 \over \left[ 1-(1-a P^2)\sigma\right]^{\epsilon}} 
+{1 \over \left[ 1-(1-a Q^2)\sigma\right]^{\epsilon}}\right.\nonumber \\
&\qquad \left.
-{1 \over \left[ 1-(1-a s)\sigma\right]^{\epsilon}}
-{1 \over \left[ 1-(1-a t)\sigma\right]^{\epsilon}}\right\}\ ,
\end{align}
where $a=u/(P^2Q^2-st)$. Now each of the four terms by itself is divergent (even for $\eps \neq 0$), only the sum gives
a finite integral. A straightforward way of regulating each term individually
is to simply subtract $1/\sigma$ from each term  in the integrand, thus
removing the 
divergence at $\sigma=0$. We thus have
\beq
  {\cal F}_\epsilon(s,t,P^2,Q^2)\ = \ {(-a)^\epsilon \over 2} \Big[ f(1-a P^2)
    +f(1-a Q^2) - f(1-as)-f(1-at) \Big] \ , 
    \eeq
    where 
    \beq
f(x)\ =  \ {1 \over \epsilon}\int_0^1 d\sigma {(1-x \sigma)^{-\epsilon}-1\over
  \sigma}={1 \over \epsilon}\int_0^x d\sigma
{(1-\sigma)^{-\epsilon}-1\over \sigma} \ . 
\eeq
The problem becomes that of finding the integral $f(x)$. It has two equivalent forms, both given in
terms of hypergeometric functions. The first form is given by
\begin{align}
\label{fexp}
  f(x)\, =\, x \times  {}_3 {\rm F}_2(1,1,1+\epsilon;2,2;x)=
  \sum_{n=1}^\infty \epsilon^n S_{1\,n+1}(x)\ .
\end{align}
Notice the very simple expansion in terms of Nielsen polylogarithms. 
The second form is
\begin{align}\label{3f2ex}
f(x) \, = \, -{1\over \epsilon^2} \left[(-x)^{-\epsilon}   {}_2
{\rm F}_1(\epsilon,\epsilon;1+\epsilon;1/x)+\epsilon \log x \right]+ \mathrm{constant}\ , 
\end{align}
where the constant is there to make $f(0)=0$, and is not important
since it will cancel in $ {\cal F}_\epsilon$.

We thus arrive at two different forms for the Wilson loop diagram. The
first form is  
 \begin{align}
 \label{fin1}
&  {\cal F}_\epsilon(s,t,P^2,Q^2)\nonumber\\
=& {(-a)^\epsilon\over 2} \Big[ (1-aP^2)   {}_3
  {\rm F}_2(1,1,1+\epsilon;2,2;1-aP^2)+(1-aQ^2)   {}_3 {\rm F}_2(1,1,1+\epsilon;2,2;1-aQ^2)
  \nonumber\\
& \qquad -(1-as)   {}_3 {\rm F}_2(1,1,1+\epsilon;2,2;1-as)-(1-at)   {}_3 {\rm F}_2(1,1,1+\epsilon;2,2;1-at)\Big] \ , 
\end{align}
and it is manifestly finite. Furthermore, since
${}_3
{\rm F}_2(1,1,1+\epsilon;2,2;x)={\text{Li}_2(x)}/{x}$, this form directly leads to the expression derived in  
\cite{Duplancic:2000sk,Brandhuber:2004yw}
for the finite $2me$ box function, 
\begin{align}
 {\cal F}_{\epsilon=0}(s,t,P^2,Q^2)
\ = \ {1\over 2}\left[ \text{Li}_2(1-aP^2)+\text{Li}_2(1-aQ^2) 
-\text{Li}_2(1-as) -\text{Li}_2(1-at)\right] \ . 
\end{align}
We also notice that the simple expansion of~(\ref{fexp}) gives a correspondingly  simple
expansion for the Wilson loop diagram in terms of Nielsen
polylogarithms.

The more familiar looking second form for the two-mass easy box function is  (see (A.13) of \cite{Brandhuber:2005kd})
 \begin{align} 
 &{\cal F}_\epsilon(s,t,P^2,Q^2) \ = -{1 \over 2 \epsilon^2} \times\nonumber \\
&
  \bigg[ 
  \left({a\over 1-aP^2}\right)^{\epsilon} {}_2
    {\rm F}_1(\epsilon,\epsilon;1+\epsilon;1/(1-aP^2)) + \left({a\over 1-aQ^2}\right)^{\epsilon} {}_2
    {\rm F}_1(\epsilon,\epsilon;1+\epsilon;1/(1-aQ^2))\nonumber\\
& -\left({a\over 1-as}\right)^{\epsilon} {}_2
    {\rm F}_1(\epsilon,\epsilon;1+\epsilon;1/(1-as))-\left({a\over 1-at}\right)^{\epsilon} {}_2
    {\rm F}_1(\epsilon,\epsilon;1+\epsilon;1/(1-at))\nonumber\\
&\qquad +\epsilon (-a)^{\epsilon}\Big(\log(1-aP^2)+\log(1-aQ^2)-\log(1-as)-\log(1-at)\Big)
\bigg]\ .
\label{fin2}
\end{align}
This second form was derived in~\cite{Brandhuber:2005kd,bht} except for
the last line which 
is an additional correction term needed to obtain the correct analytic continuation in all regimes. 
The identity
\begin{align}
  {(1-aP^2)(1-aQ^2)\over (1-as)(1-at)}\ =\ 1
  \ , 
\end{align}
implies that if all the arguments of the
logs are positive then this additional term vanishes,  but for example if we have
$1-aP^2,1-aQ^2>0$ and $1-as,1-at<0$ then the additional term gives
(taking care of the appropriate analytic continuation in (\ref{an}))
${\rm{sgn}} (a)2\pi i (-a)^\epsilon/\epsilon$. This becomes important when
considering this expression at four and five points in the Euclidean regime.

For applications in this paper we are interested in taking either one
massive leg massless (for the five-point case) or both massive legs
massless (for the four-point case).
Using the first expression for the finite Wilson loop diagram in terms
of ${}_3{\rm F}_2$ functions and using that ${}_3
{\rm F}_2(1,1,1+\epsilon;2,2;1)=\frac{-\psi(1-\epsilon)-\gamma }{\epsilon} =
-\frac{H_{-\epsilon}}{\epsilon}$ (where $\psi(x)$ is the digamma function,
$\gamma$ is Euler's constant and
$H_n$ is the harmonic number of $n$), we obtain the four- and five-point one-loop Wilson loop
expressions of \eqref{4pnt1loop} and \eqref{5pnt1loop}.

For completeness we also consider the limit with $P^2=Q^2=0$ using the
second expression for the finite diagram~(\ref{fin2}), since this and
similar expressions  have been used throughout appendix A.
When $P^2=Q^2=0$, we have $a=1/s+1/t$,  $1-as=-s/t$ and $1-at=-t/s$ and using 
\begin{align} {}_2 {\rm F}_1(\epsilon,\epsilon;1+\epsilon;1)=\epsilon
  \pi \csc(\epsilon\pi) 
\ , 
\end{align}
we get
\begin{align} &{\cal F}_\epsilon(s,t,0,0)
=-{1 \over 2 \epsilon^2} \times \nonumber \\
& \bigg[ - \left({u\over s t}\right)^{\epsilon} \left({t\over
       s}\right)^{\epsilon}  {}_2
    {\rm F}_1(\epsilon,\epsilon;1+\epsilon;-t/s)-\left({u \over s t}\right)^{\epsilon} \left({s\over
       t}\right)^{\epsilon} {}_2
    {\rm F}_1(\epsilon,\epsilon;1+\epsilon;-s/t)\nonumber\\
& +2({a})^{\epsilon} 
    \epsilon \pi \csc(\epsilon\pi) +\epsilon (-a)^{\epsilon}\Big(-\log(1-as)-\log(1-at)\Big)
\bigg]\ .
\end{align}
We wish to know this in the Euclidean regime in which $s,t,a<0$.
 The first line is then manifestly real, whereas the second gives
 \begin{align}
   2\pi\epsilon({-a})^{\epsilon}e^{-i\epsilon \pi} \csc(\epsilon\pi) +
   2\pi \epsilon i (-a)^\epsilon
   =2\pi\epsilon({-a})^{\epsilon}\cot(\epsilon\pi)\ .
 \end{align}
This is the form used for the one-loop Wilson loop throughout the
paper, for example in~(\ref{4pnt1loop})  and~\eqref{fcross}.

\end{document}